\documentclass[granma]{svjour}
\usepackage{graphics,epsfig,citesort}

\begin{document}

\title{How to handle the inelastic collapse\\
       of a dissipative hard-sphere gas with the TC model}

\author{Stefan Luding$^{(1)}$ and Sean McNamara$^{(1,2)}$}
\institute{(1)          
           Institute for Computer Applications 1, \\
           Pfaffenwaldring 27, \\
           70569 Stuttgart, GERMANY\\ 
           e-mail: lui@ica1.uni-stuttgart.de\\
           (2)            
           Levich Institute, \\
           Steinman Hall T-1M, \\
           140th St and Convent Ave,\\
           New York, NY 10031, USA\\
           e-mail: mcnamara@levdec.engr.ccny.cuny.edu\\
           ~\\
\noindent
We thank Timo Aspelmeier, Jean Rajchenbach, Stefan Schwarzer, and
Annette Zippelius for helpful discussions and acknowledge the
support of the ``Alexander von Hum\-boldt-Stiftung'' and of
the Deutsche Forschungsgemeinschaft, Sonderforschungsbereich 382.
}
\date{Received: June 29, 1998}
\maketitle

\def\n{\mathbf{\hat{n}}}
\def\bv{\mathbf{v}}

{\bf \noindent Abstract~}
The inelastic hard sphere model of granular material is
simple, easily accessible to theory and simulation, and
captures much of the physics of granular media. It has
three drawbacks, all related to the approximation that
collisions are instantaneous: 1) The number of collisions
per unit time can diverge, i.e.~the ``inelastic collapse''
can occur. 2) All interactions are binary, multiparticle
contacts cannot occur and 3) no static limit exists.
We extend the inelastic hard sphere model by defining a
duration of contact $t_c$ such that dissipation is
allowed only if the time between contacts is larger than $t_c$.  
We name this generalized model the {\em TC model} and discuss it using 
examples of dynamic and static systems. 
The contact duration used here does {\em not} 
change the instantaneous nature of the hard sphere contacts, 
but accounts for a reduced dissipation during ``multiparticle contacts''.
Kinetic and elastic energies are defined as well as forces and stresses
in the system.  Finally, we present event-driven numerical simulations 
of situations far beyond the inelastic collapse, possible only with
the TC model. 

\tableofcontents
\section{Introduction}

Granular media consist of discrete particles. Their interaction is
governed by two major concepts:  excluded volume and dissipation.  Since
the particles are solid, each particle occupies a certain amount of
space, and no other particle may enter this volume.  If another particle
approaches, the pair eventually collides.  During collisions, energy is
lost from those degrees of freedom (linear or rotational motion) which
are important for the behavior of the material.  Heat or sound are
radiated and plastic deformation takes place so that energy is
irreversibly lost \cite{herrmann98}.

A model accounting for both excluded volume and dissipation is the 
inelastic hard-sphere (IHS) molecular dynamics with dissipative 
binary interactions.
It is frequently used for the simulation of granular media, and
is attractively simple. Between collisions, particles move freely through
space.  When two particles touch, their velocities are instantly replaced
by new velocities calculated from a {\em collision rule}:
\begin{equation}
{ \cal U' = C(U,R), }
\label{eq:abstract}
\end{equation}
where ${\cal U}$ are the particles' velocities before the collision, and
${\cal U'}$ are those after the collision.  ${\cal R}$ denotes the
particles' positions at the time of collision.  In theory, ${\cal C}$
could be anything, but in practice it is restricted by physical
considerations, i.e.~the particles may not interpenetrate, Galilean
invariance, conservation of momentum, dissipation of energy, etc..
${\cal U}$ may also contain angular velocities, and ${\cal C}$ can be
chosen to mimic real particles. Simulations using the hard sphere model can 
be very fast because to simulate a collision, the computer needs only to
evaluate Eq.~(\ref{eq:abstract}).  On the other hand, if forces between
particles are specified, the computer must integrate a differential
equation over several time steps for each collision.
Note that the assumption of an ideally hard potential is also
used in kinetic theory and the Boltzmann or Enskog approaches 
\cite{haff83,jenkins85b,lun86,goldshtein95,goldhirsch98},
which facilitates comparisons between simulation and theory.
All the collision rules we consider in this paper can be written
\begin{equation}
\bv'_{1,2} = \bv_{1,2} \pm \frac{1+r}{2}
         \left[\left(\bv_2-\bv_1\right) \cdot \n \right] \n,
\label{eq:practical}
\end{equation}
where $\n$ is a unit vector joining the line of centers, and $\bv_i$ is
the velocity of particle $i$.  But the most important symbol appearing
in Eq.~(\ref{eq:practical}) is $r$, the {\em restitution coefficient}. 
The following equation for $r$ can be derived from Eq.~(\ref{eq:practical}):
\begin{equation}
r = - \frac{\left(\bv'_2-\bv'_1\right) \cdot \n}
		{\left(\bv_2-\bv_1\right) \cdot \n},
\end{equation}
so $r$ is the ratio of the component of the relative velocity along
the line of centers after the collision to its value before the collision.
If $r=1$, collisions conserve energy, and are said to be elastic.  For
$0 \le r<1$, energy is dissipated, and the collisions are inelastic.
Usually $r$ is considered to be a property of the material, and set to
a constant which is the same for all collisions.  

The IHS model is best accessible to simulations and theory, but it
has three general problems.  First, there is the singularity of inelastic
collapse: an infinite number of collisions can occur in finite time. 
Secondly, the collision rule treats only binary interactions, but
in reality, many grains can interact, and these multiparticle
interactions are different from a sequence of binary collisions.  Finally,
there are no enduring contacts between particles, and no analog to various
physical quantities, such as the energy stored in inter-particle contacts,
exists.
In this paper, we present the ``TC model'', which is an extension of the
IHS model that remedies these three problems.  The collisions are still
instantaneous, but we suppose that two particles influence each other 
during a time $t_c$ after the collision.  Specifically, if a particle
experiences two collisions separated by a time less than $t_c$, a
multiparticle event is assumed to occur, and the second collision dissipates
no energy.  Except for this additional rule, the TC model is identical
to the IHS model.

The problems of the IHS model are discussed in more detail in Sec.~II. 
In Sec.~III,
we review various attempts to solve these problems, including the TC
model.  Sec.~IV describes the TC model in detail and applies it to the
simple example of a particle lying on a flat surface.  The elastic
energy of a two-dimensional hard sphere gas is defined in section
V and in section VI results on dissipative systems are presented that
could be achieved by using the TC model whose consequences and
future perspectives are discussed in section VII.

\section{Problems of the IHS model}

In this chapter, we discuss the problems of the IHS model that any
improvement of it would have to correct.  All three problems have
essentially one origin:  the potential used between the centers of mass of
two colliding particles is unphysically stiff.  The instantaneous
collisions imply an interaction potential which is constant when there is
no contact, and suddenly becomes infinite when the particles touch.
Therefore, momentum exchange takes place in zero time and thus the
corresponding forces are infinite, however, acting for zero time only.  In
a real system the situation is different:  each contact takes a finite time
during which large, but finite forces act.  The infinitely stiff
hard-sphere interaction is only an idealization or simplification of a
smooth repulsive pair-potential.

\subsection{Inelastic Collapse}

The most dramatic consequence of the infinitely stiff interaction potential
used in the IHS model is inelastic collapse, which manifests itself
as an infinite number of collisions in finite time.  It was first
discovered while studying the one-dimensional (1D) model system 
of a column of dissipative particles hitting a wall. 
The occurence of the inelastic collapse can be estimated
using the product of the number of particles $N$ and the dissipation
per contact $(1-r)$.  The effective dissipation $\xi = N (1-r)$
has a critical value of $\xi_c \approx \pi$ above which collapse
occurs. The above value of $\xi_c$ was calculated
with the independent collision wave (ICW) model \cite{bernu90}.
With slightly different arguments using the ``cushion model'' 
\cite{mcnamara92}, the value was evaluated as 
$\xi_c \approx \ln[4/(1-r)]$. The ICW model seems to work
better in the inelastic limit, whereas the cushion model is superior
in the elastic limit \cite{mcnamara92}.

Inelastic collapse is also present in two dimensions (2D).  In freely
cooling systems a minimum of three particles is enough to lead to the
collapse, if dissipation and density are large enough 
\cite{mcnamara94,goldman98}.
In larger assemblies, the inelastic collapse occurs, but it involves just a
few particles arranged almost along a line.  This leads to the conclusion
that the inelastic collapse is mainly a 1D effect \cite{esipov97} and that
the one-dimensional predictions for the critical $\xi$ should work also in
2D.  In fact, inelastic collapse
in 2D unforced simulations can be predicted reasonably well by using the
1D criteria with $\xi = (\nu l/d)(1-r)$ ($\nu$ is the fraction of the
total area covered by the disks, $l$ is the lenght of one side of the
domain, and $d$ is the disk diameter) \cite{mcnamara96}.
In a container in the presence of gravity, this expression for $\xi$
is equivalent to the number of layers of particles when the granular
material is at rest.
With these boundary
conditions, the inelastic collapse likely occurs for small energy input and
large $\xi$ \cite{grossman97b}.  Vibrated containers with large filling
heights cannot be simulated with the IHS model
\cite{luding96e,grossman97b}.

In two dimensions only a small fraction of 
the particles in the system is involved in the inelastic collapse. 
This implies that it
is both physically insignificant for real particle assemblies
and a major drawback for numerical simulations of dissipative
systems.  Therefore, any improvement of the IHS model must avoid the
singularity of inelastic collapse.

\subsection{Multiparticle interactions}

In the IHS model, true multiparticle interactions are impossible because
collisions are instantaneous.  Multipar\-ticle interaction is
built up out of many two-particle interactions.  In
a real system the situation is different:  Each contact takes a finite time
so that multiparticle contacts are possible.
The difference between two- and multiparticle contacts was examined for
one- and two-dimensional model systems \cite{luding94d,luding95,luding94b},
and it was found that {\em less} energy is dissipated in multiparticle
events than in an equivalent sequence of binary collisions.  Any improvement
of the IHS model should have the property that energy dissipation is
reduced during multiparticle interactions.

\subsection{No static limit}
\label{sec:static_limit}

Another problem with the IHS model is that the static limit does not
exist, i.e.~there is no way to represent enduring contacts between
particles. For example, in the framework of the IHS model, a particle
cannot rest motionless on the ground. We discuss this simple example
in more detail in subsection\ \ref{sec:bounce}. Another way to formulate
this problem is by considering the energies in the system, see also
subsection\ \ref{sec:energies}, and especially the elastic contact 
energy that is not defined in the IHS model. 

The translational kinetic energy $E$ of the system is 
$E=(1/2)\sum_{i=1}^N m_i (v_i^2+u_i^2)$, with the mass $m_i$, the 
fluctuation velocity $v_i$, and the flux velocity $u_i$ of particle 
$i$. In the following we will mainly consider situations with $u_i=0$. 
The potential energy $E_p$ 
is zero in absence of an external body force like e.g.~gravity.
In addition to $E$ and $E_p$, real materials have an elastic 
energy $E_{el}$ at each contact, which is {\em not} defined in the context
of the IHS model. In classical elastic systems, the total
energy, i.e.~the sum of kinetic, potential, and elastic energy,
$E + E_p + E_{el} = E_{\rm tot}$, is always conserved. 
In a dissipative system without a source of energy, the total energy 
tends towards a constant while the kinetic energy tends towards zero
in the long time limit. This state is referred to as static,
not to be confused with a ``quasi-static'' state, defined in the 
context of the TC model below in subsection \ \ref{sec:energies}.\\
More specifically, consider a realistic granular material inside a box,
under the influence of gravity and in the absence of energy sources. 
Due to dissipation, the material will loose
energy. The potential and elastic energies will approach
constants values while the kinetic energy tends towards zero. 
Eventually all particles are at rest, with $E = 0$, and the elastic 
energy is, as a rule, larger than zero, since a certain number
of contacts is necessary to allow for a stable static configuration. 
If the particles would be dissipative hard spheres, the inelastic
collapse can occur long before $E$ vanishes so that the system will 
never reach a configuration with constant $E_{\rm tot}$. The inelastic
collapse brings the system to an artificial halt.
A static configuration with zero kinetic energy could be reached 
in a hard-sphere system by piling the spheres on top of each other
(just in contact with no overlap). For this situation $E_{el}$ is
not well defined and those touching hard spheres violate the 
rule of instantaneous contacts. Since in this artificial limit 
no elastic energy is defined, contact-forces and stresses are also 
not properly defined.  This is an argument against the IHS model
itself, which has by construction no static or ``zero-temperature''
limit.\\
A method that loosens the restriction of instantaneous contacts 
and in return defines contact forces is the so-called contact dynamics, 
(see Ref.\ \cite{radjai98} and references therein).  The TC model, which
we present in this paper, also provides a way to define the elastic energy.

\section{Proposed modifications to the IHS model}
\label{sec:avoid}

In this section we review the various extensions and modifications of the
IHS model which have been proposed.  The main goal of these suggestions 
has been to remove inelastic collapse, since it is the most conspicuous
problem of the IHS model.

\renewcommand{\theenumi}{~(\roman{enumi})}
\begin{enumerate}

\item 
\label{i:lrv} 
Particles with relative energy below a critical threshold can
be merged into a ``cluster'' by setting their relative velocity
and separation to zero. If another particle hits such a cluster,
the momentum transport inside the cluster takes place instantaneously
in the sequence of the largest relative velocity (LRV) \cite{luding94}.
With this method clusters can grow, and, given strong enough energy
input, clusters may also be destroyed again. The 
deterministic LRV model has been successfully implemented in 1D,
but stresses and energies in the bulk are not defined.\\

\item 
\label{i:stochastic} 
Several authors suggest a stochastic addition of translational or
rotational energy, as soon as the relative velocity after a collision
drops below a critical value \cite{deltour97,grossman97b}. Also a 
small rotation of the relative velocity after contact of less than 
5 degrees, seems to hinder the inelastic collapse \cite{deltour97},
since correlations between successive collisions are diminished.
Those stochastic models prevent the occurence of the inelastic 
collapse in the systems examined. However, neither is it clear
whether the collapse can be circumvented under all conditions, 
nor has the physical relevance of the random energy input been discussed
so far.\\

\item
\label{i:modes}
Another method involves internal modes of every particle.
At each collision these modes may be agitated, and their energy
is dissipated only on a time-scale longer than the
duration of a contact. If a particle suffers an additional
collision within this time, then energy can be transferred
from the internal modes back into translational motion.
At least in a cooling system in 1D the inelastic collapse is
prevented \cite{giese96,aspelmeier98}, and possibly this
model leads to an explanation of the random energy input
mentioned above \ref{i:stochastic}. The drawback of this
method is the effort necessary to model the internal 
modes.\\
 
\item 
\label{i:rv}
A frequently used way to reduce dissipation in the low velocity regime
(which has also been observed experimentally)
is a velocity dependent restitution
coefficient $r(v)$ based on the assumption of either viscoelastic
\cite{luding94d,brilliantov96} or plastic \cite{walton95b,thornton97b}
contacts.  The velocity dependence seems to avoid the inelastic collapse in
the absence of walls and external forces \cite{goldman98}, however, its use
for other boundary conditions resembling systems on earth, has not been
examined up to now.  Note that the dependence of $r(v)$ on the collision
velocity still concerns binary collisions with varying velocity, whereas 
the TC model discussed below concerns the transition from binary to 
multiparticle contacts -- a completely different approach.  
Even when a velocity dependence of $r$ can simply be added to
the TC model, we avoid this in the following for the sake of simplicity.\\

\item 
\label{i:TC}
Instead of feeding energy into the system, another approach
just switches off dissipation following certain rules. 
The idea is to decide whether a particle still feels its previous
collision partner when colliding with the next one. This is 
important since the presence of a third particle will affect 
the collision of a given pair.

One can switch off dissipation if the next collision partner of a particle 
is detected within a critical distance $\lambda_c$ \cite{mcnamara96}.
In fact, it makes no sense to treat particles as separate objects if
their surface-surface distance is of atomic size and, more technically, 
the numeric errors can become large when the distance between
the particles is many orders of magnitude smaller than the particle
diameter. 

A similar argument for switching off dissipation is
that two collisions cannot be treated as separate events if 
they take place within a too short time-interval $t_c$
that corresponds to the duration of a contact. It has been shown 
that collisions which overlap (in time) lead to weaker dissipation 
than a series of binary collisions (separated in time) 
\cite{luding94d,luding95,falcon98b}. 
Thus one sets the restitution coefficient to its elastic limit 
$r=1$, if a particle suffers 
more than one collision within time $t_c$ \cite{luding96e,luding97c,luding98e}.
With the time $t^{(i)}_n$, that passed by since the last collision $n-1$ of
particle $i$, the restitution coefficient for its $n$-th collision can be 
expressed as
\begin{equation}
r^{(i)}_n = {\left \{
             { \begin{array}{lll}
                 {r} & {\rm ~~for~ } & t^{(i)}_n >  t_c\\
                 {1} & {\rm ~~for~ } & t^{(i)}_n \le t_c ~ ,
               \end{array} }
           \right . }
\label{eq:epsnew}
\end{equation}
with $0 < r \le 1$. Thus the type of a collision changes from 
inelastic to elastic when collisions occur too frequently. 
More specifically, a collision is elastic if {\em at least one partner}
fulfills the above condition $t^{(i)}_n \le t_c$.
In general, $r$ {\em and} $t_c$ can depend on the relative velocity
and other parameters
\cite{luding94d,brilliantov96}, so that the material's behavior can 
be adjusted appropriately using this dependence as discussed above 
\ref{i:rv}.  Since Eq.\ (\ref{eq:epsnew}) involves the duration of a 
contact $t_c$ we refer to this model as TC model in the following.

Before turning to some more details of the TC model, we should
mention that by using a velocity dependence for either the
dissipation cut-off distance $\lambda_c$ or the dissipation cut-off
time $t_c$, both can be connected:
Assuming $t_c(v) \propto v^{-\alpha}$, with the typical relative
velocity $v$, leads to $\lambda_c \propto v t_c(v) \propto v^{1-\alpha}$.
With $\alpha=1$ one has $\lambda_c = {\rm const}$, and alternatively, 
with $\lambda_c \propto v$ one gets 
$t_c \propto \lambda_c / v = {\rm const}$..
This unifies both the critical time and the critical length criteria 
into a single framework. In the following sections we will focus on the 
cut-off time (or contact duration) $t_c$.\\

\end{enumerate}

To our knowledge, none of the models mentioned above has a solid 
theoretical background, except for the approach in \ref{i:modes} 
involving internal modes.
The LRV method \ref{i:lrv} has no reasonable static limit
where stresses can be defined. The stochastic approaches
\ref{i:stochastic} require the choice of an a priori unknown
source of fluctuation energy. 
The velocity dependence of $r$ in \ref{i:rv} was experimentally
measured for binary collisions, and is not necessarily important
for multiparticle contacts. In the following we will discuss the physical
relevance of the TC model, define its quasi-static limit, and finally
apply it to selected examples.

\section{The TC model in detail}

\begin{figure}[ht]
\begin{center}
\epsfig{file=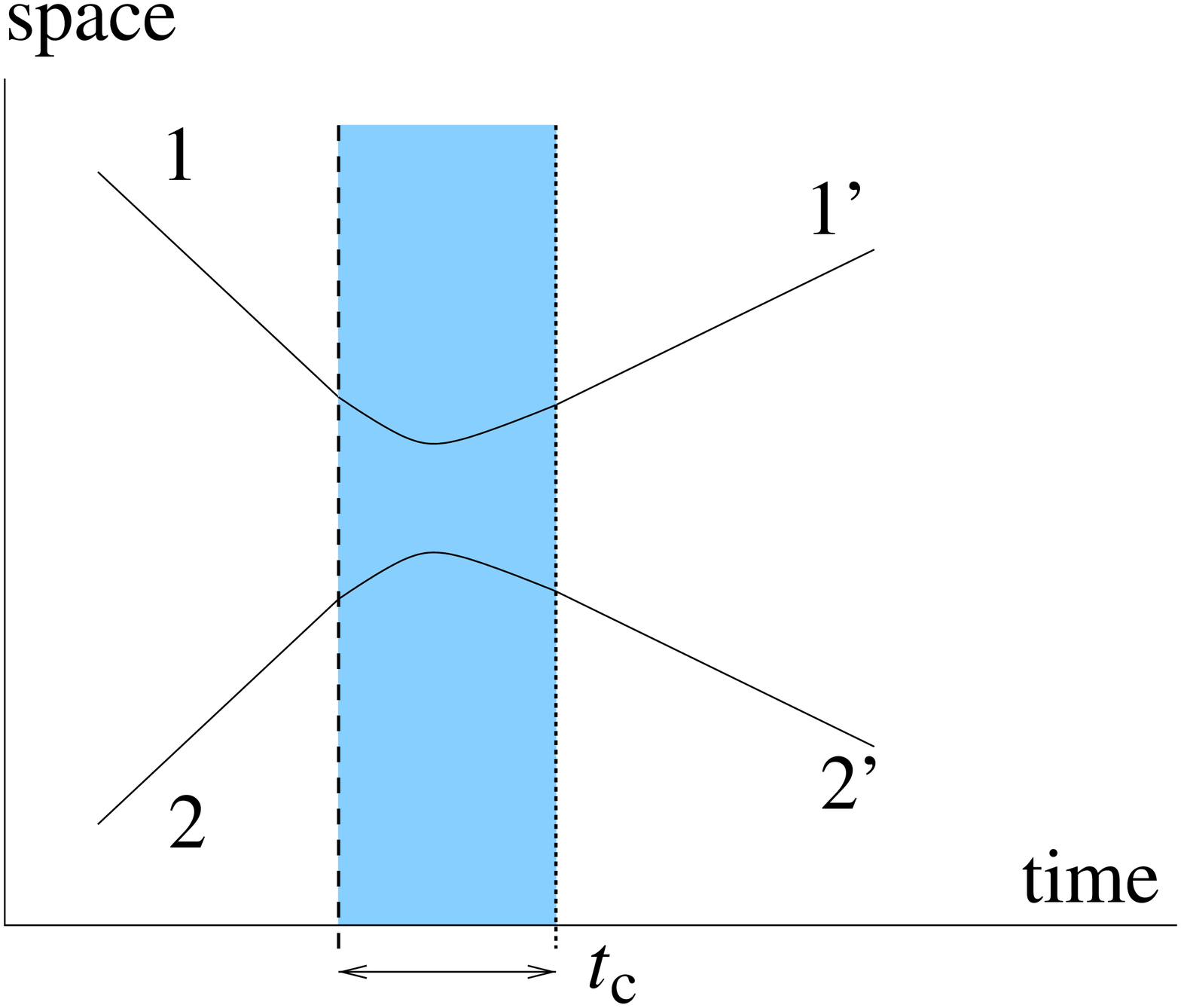,height=3.5cm,angle=0} 
\epsfig{file=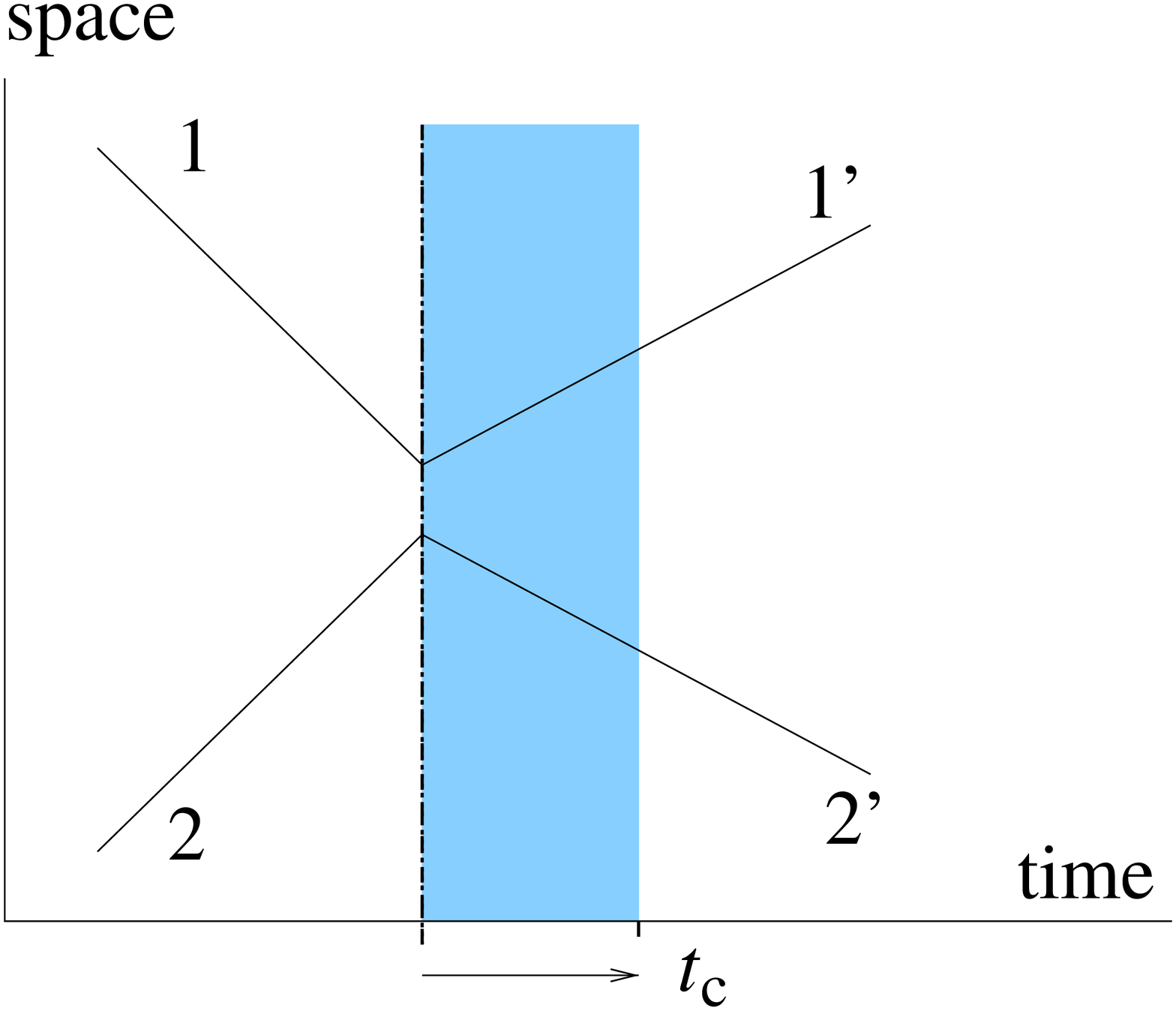,height=3.5cm,angle=0} 
\end{center}
\caption{Schematic plot of the trajectories of two soft (left) and
two hard (right) particles against time. The beginning and the ending 
of the interaction are marked by dashed and dotted vertical lines respectively
and the time $t_c$ during which dissipation is affected is marked as shaded 
region.
}
\label{fig:schem1}
\end{figure}

In Fig.\ \ref{fig:schem1}, an interaction of two `soft' particles
(left) is compared to the interaction of two `hard' particles (right).
The words `hard' and `soft' correspond to non-smooth and smooth potentials
between the centers of mass of two particles respectively, and both are
different approaches to model solid body interaction.
Soft particles interact essentially during a time $t_c$, indicated
by the shaded region and the dashed vertical lines which mark the beginning
and the ending of the contact. The contact duration $t_c$ of the soft
particles is defined by these two instants of time. In the case of the hard
particles, the interaction is instantaneous and the beginning and the
ending coincide. However, in the TC model, the particles are considered 
to influence each other also during a time $t_c$ after each collision 
(the shaded region). Note that the TC model in the limit $t_c=0$ is 
identical to the IHS model and that both are identical to the elastic 
hard sphere model when $r=1$.

Assuming that $t_c$ is the physical duration of a contact, 
multiparticle contacts take place whenever a further collision
of the particles in Fig.\ \ref{fig:schem1}(b) occurs within
the shaded area. This is the case when the typical 
time between instantaneous contacts $t_n$
gets smaller than the duration of a contact $t_c$. The ratio 
$\tau_c = t_n/t_c$ is a measure for the existence of multiparticle
contacts \cite{luding94d}. If $\tau_c \gg 1$ pair interactions dominate, 
whereas for $\tau_c \ll 1$, one particle may be in contact with several
others. Numerical simulations with various soft interaction potentials
show that dissipation gets more and more ineffective with 
decreasing $\tau_c$ \cite{luding94d,luding95}.

\subsection{The energies in the system}
\label{sec:energies}

In this subsection, we show how an elastic energy can be defined in the TC
model.  As was discussed above, there is no elastic energy in the IHS model. 
In a real system, only particles in contact with each other
contribute to the elastic energy.  As a consequence, we consider all
particles which collided no longer than $t_c$ ago to be still in contact,
and thus their energy contributes to a pseudo-elastic energy $E_e$.  The
translational kinetic energy $E$ splits into a free kinetic energy $E_k$
and the pseudo-elastic energy $E_e$ so that $E_k = E - E_e$.  Free means
that the kinetic energy $E_k$ can be dissipated whereas elastic energy
$E_e$ cannot.  In both the IHS and TC models, $E_{el}$ does not exist due
to the non-smooth interaction potential, but in the TC model, it is
replaced by $E_e$.  In table \ref{tab:energies} we summarize the meaning of
the symbols in the framework of the different models.
From table \ref{tab:energies} it is obvious that some type of
elastic energy is missing in the IHS model, while the TC model
has three types of energy like a real system. However, a direct
quantitative comparison of the energies in the framework of
a realistic system to those in the TC model is far from the scope
of this study \cite{hansen86}.

\begin{table}[ht]
\begin{center}
\begin{tabular}{ l c c c }
\hline
                       & Real        & IHS        & TC        \\
\hline
\hline
(free) kinetic energy  & $E$         & $E$        & $E_k$  \\
\hline
contact energy         & $E_{el}$    & 0          & 0      \\
elastic, kinetic energy& 0           & 0          & $E_e$  \\
\hline
potential energy       & $E_p$       & $E_p$      & $E_p$  \\
\hline
\end{tabular}
\end{center}
\caption{Meaning of the symbols $E$, $E_k$, $E_{el}$, $E_e$, and 
$E_p$ in the framework of a real system (or soft particle model), the
classical hard sphere model (IHS) or the TC model.}
\label{tab:energies}
\end{table}
As already mentioned in subsection\ \ref{sec:static_limit} the static
limit of a real system is reached when $E=0$, and $E_{el}+E_p=$const..
The corresponding state of the TC model is, consequently, identified
by  $E_k=0$, and $E_{e}+E_p=$const.. However, we denote it as
the ``quasi-static'' limit, because  $E_{e} > 0$ implies translational motion
due to the kinetic nature of the elastic energy $E_e$. 
A similar situation with $E=0$, and
$E_p=$const. in the framework of the IHS model would mean that
the system is artificially frozen. \\

For example, consider a periodic system with realistic dissipative particles 
and without external forces ($E_p = 0$) in its center of mass reference 
frame. Starting from an initial configuration with $E>0$ the system will
evolve in time and the energy will decay. Note that this boundary conditon
is totally different from the situation discussed in section\ 
\ref{sec:static_limit} 
when gravity was active and walls were present. The only stable static 
situation (in the sense that $E_{\rm tot}=$const.) would be one with all 
particles at rest ($E=0$) and isolated, possibly just touching each other 
($E_{el}=0$). Such a situation one can denote as {\em artificially frozen}.
Only in this case, kinetic and elastic energy vanish and the 
total energy can remain a constant. If two particles would touch
each other with $E_{el}>0$, at least a part of their elastic energy will be
transferred into relative velocity, eventually separating them,
so that $E>0$. This motion would eventually lead to more collisions
reducing the total energy further. In other words:
since no attractive or external forces are involved, there is no reason for
a stable overlap or deformation to exist in the long time limit. 
A similar argumentation for
the TC model leads to the analogous conclusion $E_k = E_e = 0$.
In all other situations the system evolves with time and the total energy
is not constant. The IHS model, in contrast, comes to a halt when
the inelastic collapse occurs and the long time limit cannot be reached.
 
\subsection{The special case of one bouncing particle}
\label{sec:bounce}

\begin{figure}[ht]
\begin{center}
\epsfig{file=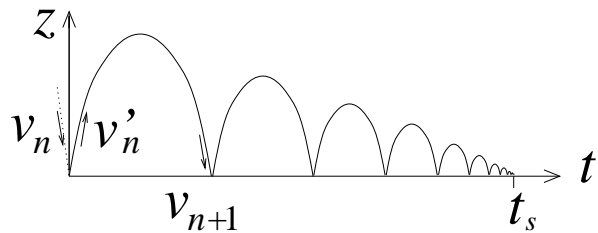,height=3.4cm}
\end{center}
\caption{Trajectory of a bouncing particle on a flat surface
as a function of time. At time $t_s$ the particle is at rest.
}
\label{fig:bounce}
\end{figure}

In order to discuss forces and stresses in the quasi-static
limit, we examine the simplest possible case of a ball
bouncing on a flat plane under the influence of the
gravitational acceleration $g$ pointing in negative $z$
direction. This system is essentially 1D, and
in the elastic limit, the particle
will bounce forever. Introducing dissipation, with 
the restitution coefficient $r$, leads to a 
velocity just after the $n$-th collision $v'_n = -r v_n$ 
as a function of the velocity $v_n$ just before. 
With the initial velocity $v_1$, one has
\begin{equation}
v_{n} = r^{n-1} v_1~.
\end{equation}
The time between the collisions $n$ and $n+1$ is 
\begin{equation}
t_{n+1} = 2 v_{n+1} / g~,
\label{eq:tn+1}
\end{equation}
for negative $v_{n+1}$ and negative $g$. 
In Fig.\ \ref{fig:bounce} the vertical position of a bouncing particle
is plotted schematically. At each collision energy is lost and the
particle stops at time $t = t_s$.

If the particle is infinitely rigid, as assumed by the IHS model, the
particle will bounce an infinite number of times before $t_s$.
At times greater than $t_s$, the IHS model is no longer defined.
But this picture of an infinitely rigid bouncing ball makes no sense as
soon as $t_n$ gets comparable to the duration of a contact $t_c$. In that case
the particle is in steady contact with the plate \cite{falcon98}. 
Therefore, in the TC model, $r$ is set to $1$ when $t_n<t_c$,
and the particle bounces forever on the plate with a constant
period which is less than $t_c$.  This is the TC model's representation
of a particle lying on the plate.  We now compare $t_s$ in the IHS
and TC models.
Starting with $v_1 \gg g t_c / (2 r)$, the quasi-static limit is
reached when $t_{n+1} \le t_c$ so that 
\begin{equation}
n_s \ge \frac{\log \left (\frac{g t_c}{2 v_1} \right )}{\log ~r}~.
\end{equation}
In the hard-sphere limit $n_s \rightarrow \infty$ since $t_c \rightarrow 0$.
The time until the particle has lost all its kinetic energy is
\begin{equation}
t_s^{\rm(IHS)} =
\sum_{n=1}^\infty t_{n+1} = \frac{2 v_1 r}{g} \sum_{n=0}^\infty r^n 
         = \frac{r}{1-r}t_1~,
\label{eq:ts_infty}
\end{equation}
with $t_1$ from Eq.\ (\ref{eq:tn+1}).
The time until a particle with $t_c > 0$ reaches the
quasi-static regime is
\begin{equation}
t_s^{\rm(TC)} =
\sum_{n=1}^{n_s} t_{n+1} = \frac{2 v_1 r}{g} \sum_{n=0}^{n_s-1} r^n 
                  = t_s^{\rm(IHS)} (1-t_c/t_1).
\end{equation}
The difference $\Delta t_s$ between these two times is a measure for
the difference between a soft particle ($t_c >  0$) and a hard-particle
($t_c=0$) model:
\begin{equation}
\Delta t_s = t_s^{\rm(IHS)} - t_s^{\rm(TC)} 
     \approx \frac{r}{1-r}t_c.
\label{eq:deltats}
\end{equation}
The last term in Eq.\ (\ref{eq:deltats}) is obtained by assuming that the
time between elastic collisions $n>n_s$ is approximately $t_c$ (what is
almost true for $r \approx 1$).
Note that $\Delta t_s$ will be small, because the contact time is
usually small. 

\subsubsection{Performing averages}
\label{sec:aver}

In the framework of the TC model, an observable $A$ has to be defined
in average over a time-interval $\Delta t$:
\begin{equation}
A(t) = \left < A \right >
 = \left < A\right > _{\Delta t}
 = \frac{1}{\Delta t} \int_{t-\Delta t}^{t} A(t')~{\rm d}t' .
\end{equation}
Alternatively, ensemble averages can be performed,
however, this option will not be discussed here. The average makes sense
only if it averages at least over the duration of a contact $t_c$, since the 
TC model simplifies the reality during times smaller than $t_c$. Therefore, 
averages over longer intervals should be taken to level out the details of 
the basic assumptions introduced in e.g.~Eq.\ (\ref{eq:epsnew}).

We now discuss this average, using as an example one particle 
resting on a flat surface.
A real, soft particle resting on the bottom is represented 
in the TC model as an elastic, hard particle bouncing on the 
surface with a period $t_n \le t_c$.  Since it performs a 
periodic orbit with duration $t_n$, one can set $\Delta t = t_n$.
Thus, integration is performed over one parabola
of free flight of the particle. The mean velocity of the bouncing particle
is $u = \left < v \right > = 0$, as expected for the quasi-static limit,
the mean squared fluctuation velocity is 
$\left < (v-u)^2\right > = \left < v^2\right > = (1/12)(g t_n)^2$,
and the mean separation of the particle from the bottom is
$\left < z\right > = (1/12)g t_n^2$.
With these quantities we may identify the energies defined above.
The potential energy is connected to the separation from the bottom
$E_p = m g \left < z\right > = (1/12) m(g t_n)^2$, disregarding an 
additive constant. The total translational energy is $E = E_k + E_e$, 
with  $E=(m/2) \left < v^2 \right > = (1/24) m(g t_n)^2$. 
Now, we identify the elastic energy with the kinetic energy of
the particle(s) which suffered a collision no longer than $t_c$
ago. Since the particle collides with a rate $t_n^{-1} > t_c^{-1}$, all
its kinetic energy contributes to $E_e$ so that $E_k = 0$. Note that the 
values $E_k=0$, $E_p > 0$, and $E_e > 0$ correspond to the quasi-static 
limit discussed above. 

In addition, one can calculate the force which the
particle exerts onto the bottom as the momentum exchange per
unit time $f = \left < \Delta p \right >  = m g$, as to be expected for a
particle with mass $m$ in the gravitational field \cite{luding97c}.

\subsubsection{The link to a linear elastic particle}

A soft, elastic particle in contact with the bottom has - in the 
framework of the simplest linear model \cite{luding98c} - the elastic energy
$V(\delta) = (1/2) k \delta^2$, with stiffness $k$, and overlap 
or deformation $\delta$. At rest, it exerts the force 
$f_k = k \delta_0 = m g$ onto the bottom, so that the overlap or
deformation is $\delta_0 = m g / k$. When bouncing, its
contact duration is 
\begin{equation}
t_c^{el} = \pi / \omega = \pi \sqrt{m/k}~, 
\end{equation}
what leads to the identity $\delta_0 = g (t_c^{el}/\pi)^2$.
The elastic energy of the particle at rest is thus 
$V(\delta_0) = m (g t_c^{el})^2 / (2 \pi^2)$.

Comparing the soft particle with the TC model from the previous 
subsection by using either of the relations 
$\left < z \right > \equiv \delta_0$ 
or $E_e \equiv V(\delta_0)$, leads to 
\begin{equation}
t_n \equiv \sqrt{12}\, t_c^{el} / \pi ~.
\end{equation} 
From the beginning of this section we remember that the particle
reaches its quasi-static limit when $t_n \le t_c$. Thus the
contact duration $t_c \approx t_n$
can be identified with the contact duration of the linear 
soft-sphere model $t_c^{el}$, when disregarding the constant factor
$\sqrt{12}/\pi \approx 1$. 

\subsubsection{A Gedanken-Experiment}

In order to clarify the meaning of the different energies calculated
above, we assume that we are able to switch off gravity at any time
during the period $t_n$. 
A real particle lying on a table with zero
kinetic energy will then begin to rise due to the elastic energy
stored in the contact. Ideally, for $r=1$, all elastic energy will 
be transferred to translational motion. Within the framework of the
TC model, as discussed here, one can 
calculate the velocity a particle will eventually reach, after $g$ is 
set to zero. Since the particle-velocity is phase-dependent, 
one has to perform the average over all possible phases at which 
gravity might be switched off.
Thus the integration over the period $t_n$ has to be split:
During the first half-period for $t <  t_n/2$, the particle will
keep its upwards velocity and move away from the bottom. In the second 
half-period the particle will suffer one collision with 
the bottom before it moves upwards. Performing the integration
one gets $\left < v^2_{g \rightarrow 0} \right > = (1/24) (g t_n)^2 (1+r)$. 
Note that the velocity after switching off gravity, $u_{g \rightarrow 0}$,
depends on the time when gravity is switched off. This reflects the
fact that a too fine resolution in time uncovers some details of 
the simplifications in the TC model.
However, the elastic case $r \rightarrow 1$ corresponds to the case when all 
elastic energy $E_e$ from the quasi-static regime is transferred into kinetic 
energy $E_k$.

\section{The TC model in elastic systems}

In the following we will mainly focus on the elastic limit
$r \rightarrow 1$, and define the properties of interest like
the stress tensor, the equation of state, the collision rate,
and the elastic energy. For the simulations in this section we
use an event driven (ED) simulation method as introduced by Lubachevsky
\cite{lubachevsky91}. The system has periodic boundaries, and
neither walls nor gravity are present. The following discussion 
concerns only systems in equilibrium and in their center of mass
reference frame. Strictly speaking, {\em all} collisions in the 
simulations discussed in this section are ``elastic'', because
$r=1$. However, the TC model distinguishes between those collisions
which indicate multiparticle events (where one of the partners has
had a collision no longer than $t_c$ ago), and the rest of the
collisions which are truly binary. In this section, only the former
type of collisions are called ``elastic'', for they would also
be elastic (in the TC model) when $r \ne 1$. We choose $r=1$ here
because this gives systems in equilibrium and allows to obtain
good statistics by taking long time averages.

\subsection{The stress tensor}

The stress tensor, defined for a test-volume $V_c$, can be
written component-wise as
\begin{equation}
\sigma_{\alpha \beta} = \frac{1}{V_c} 
   \left [ 
      \sum_j \ell_\alpha f_\beta - \sum_i m_i v_\alpha v_\beta
   \right ] .
\end{equation} 
The first sum runs over all contact
points $j$, and the second sum runs over all particles $i$, both
within $V_c$ \cite{goddard86,walton86}.
The indices $\alpha$ and $\beta$ denote the Cartesian coordinates,
$\ell_\alpha$ are the components of the vector from the center of mass
of a particle to its contact point $j$, where a force with components 
$f_\beta$ acts. A particle $i$ has a mass $m_i$ and a velocity
with the components $v_\alpha$. 

In the static limit, the second term drops out, since all velocities
vanish. In a dilute system without permanent contacts, the first term 
would be negligible. On the other hand, for a hard-sphere gas, the first 
term has to be treated differently, since no forces are defined. The 
dynamic equivalent to $f_\beta$ is the change of momentum per 
unit time $\Delta p_\beta / \Delta t$ \cite{luding98c}. For a hard-sphere 
gas the stress due to collisions may be evaluated as an average over
all collisions in the time interval $\Delta t$ 
\begin{eqnarray}
\nonumber
\sigma_{\alpha \beta}(t) =
\left <  \sigma_{\alpha \beta} \right > _{\Delta t} = 
~~~~~~~~~~~~~~~~~~~~~~~~~~~~~~~~~~~~~~~~\\
\frac{1}{V_c} 
   \left [
      \frac{1}{\Delta t} \sum_n \ell_\alpha(t_n) \Delta p_\beta(t_n)
      - \sum_i m_i v_\alpha(t) v_\beta(t)
   \right ] .
\end{eqnarray} 
Here, the first sum runs over all collisions $n$ 
occuring in the time between $t-\Delta t$ and $t$.
In general, the volume $V_c$ and the time-interval
$\Delta t$ have to be chosen large enough to allow averages 
over enough particles and enough collisions, but also small enough
to resolve inhomogeneities in space and variations in time. 
Note that the result may depend on the choice of $\Delta t$ 
and $V_c$ \cite{goldhirsch98}, so that this averaging procedure 
is not necessarily the best choice under all circumstances.

\subsection{The equation of state}

\begin{figure}[ht]
\begin{center}
\epsfig{file=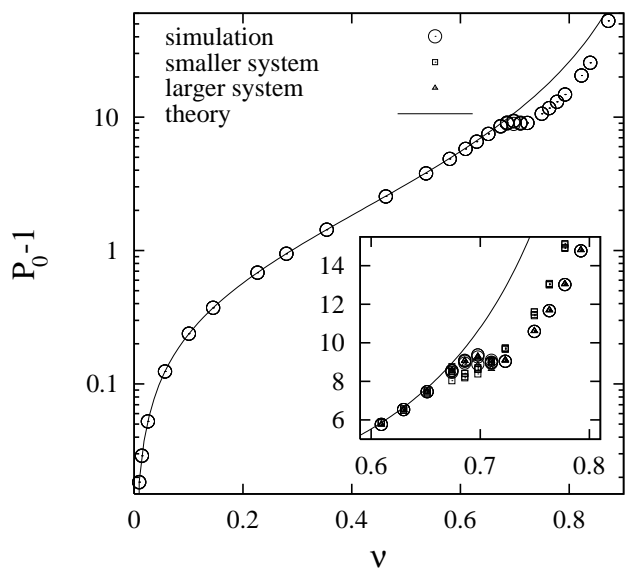,height=8.5cm,angle=-90,clip=}
\end{center}
\caption{Reduced dimensionless pressure $P_0-1$ plotted against volume
fraction $\nu$. Theory (solid line) is compared to simulations (circles) from 
a system with $N=1435$ particles. In the inset simulations of a smaller 
and a larger system, with $N=42$ (small squares) and $N=16680$ 
(small triangles), respectively, are presented in addition.
At least three simulations were performed for each $\nu$ value
and plotted over each other so that smeared out symbols indicate 
comparatively large fluctuations at $\nu \approx 0.7$.
}
\label{fig:etap}
\end{figure}

In order to test the averaged stresses defined above, we compute the mean 
pressure $P = (\sigma_1 + \sigma_2)/2~$ from the eigenvalues $\sigma_1$ 
and $\sigma_2$ of the stress tensor. The results of simulations with
elastic particles, $r=1$, and different volume fractions $\nu$ are
compared to the pressure obtained by the two-dimensional equivalent of 
the Carnahan-Starling formula \cite{jenkins85b,hansen86}. 
In Fig.\ \ref{fig:etap}, we plot the 
reduced, dimensionless pressure 
\begin{equation}
P_0 - 1 = \frac{ P V }{ E } - 1 = 2 \nu g(2a)
\label{eq:p01}
\end{equation}
against the volume fraction $\nu = N \pi a^2 / V$ \cite{hansen86}.
We use the pressure $P$, the volume of the system $V$, the total 
translational energy $E = (1/2) \Sigma_{i=1}^N m_i v_i^2$, and the 
particle-particle correlation function evaluated at contact
\begin{equation}
g(2a) = \frac{1-\frac{7}{16}\nu}{(1-\nu)^2}~, 
\end{equation}
taken from Eqs.\ (28) and (110) in Ref.\ \cite{jenkins85b}. 
The agreement between theory and simulations is perfect since $r=1$,
already for a tiny system with $N=42$.
The simulations deviate from theory only for volume fractions larger than
about 0.7 because then a solid state exists, i.e.~the motion of
the particles is hindered by their neighbors \cite{ziman79}. The smallest
system deviates slightly from the larger ones, indicating finite size
effects in the crystalline, high density regime. However, we will not
discuss the transition from a fluid, disordered to a `solid', ordered
system here.

\subsection{The collision rate}

\begin{figure}[ht]
\begin{center}
\epsfig{file=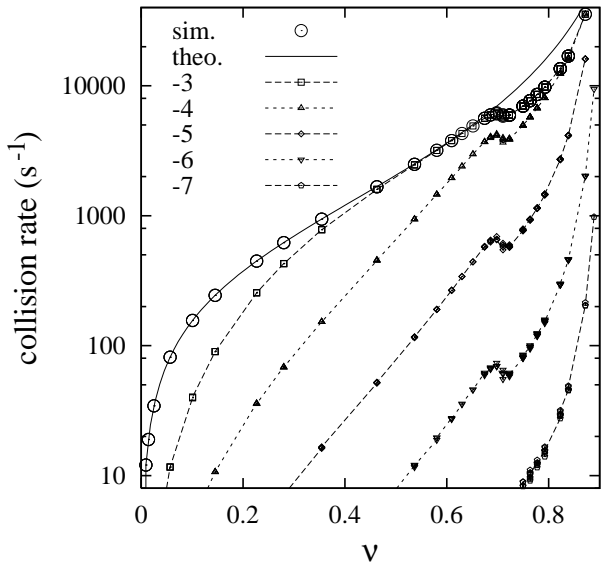,height=8.5cm,angle=-90,clip=}
\end{center}
\caption{
Collision rates plotted against volume fraction $\nu$ 
from theory (solid line) and the simulations (circles) from
Fig.\ \protect\ref{fig:etap}. The dashed lines are inserted 
between the measured elastic collision rates (small symbols)
to guide the eye. The system dependent prefactor
of these simulations is $\omega_0 = 656.03$~s$^{-1}$.
The small symbols indicate the collision rates of elastic
collisions according to the TC model with $\log_{10} t_c$
given in the inset.
}
\label{fig:tcn}
\end{figure}

The next quantity of interest is the collision rate, i.e.\ 
the number of collisions per particle per unit time. 
Thus, we define the collision rate in the simulations as the 
inverse of the typical time between contacts 
\begin{equation}
{\cal C}_r = t_n^{-1} = \frac{2 {\cal C}}{N \Delta t}~,
\label{eq:Cn}
\end{equation}
with ${\cal C} = \left < {\cal C} \right >$, the number of collisions per 
averaging time $\Delta t$. Note that we use $\cal C$ as the number of 
collisions within a time $\Delta t$, whereas the total number of collisions 
since the beginning of the simulation is denoted as $C_t$ later on. 
The prefactor `2' stems from the fact that each collision involves two 
particles. In Fig.\ \ref{fig:tcn} we compare the simulation results with the
Enskog collision frequency in 2D, as expressed in Ref.\ \cite{luding98d},
\begin{equation}
t_E^{-1} = 4 a \frac{N}{V} \sqrt{\pi \frac{E}{M}} \, g(2a)
         = \omega_0 \left ( P_0-1 \right )~,
\label{eq:tE-1}
\end{equation}
with the particle radius $a$, the total fluctuation kinetic energy $E$, 
the total mass $M$, and $g(2a)$ as defined in the previous subsection. 
From Fig.\ \ref{fig:tcn} we observe again a perfect agreement between
theory and simulation (circles) for $\nu < 0.7$.
The difference between Eqs.\ (\ref{eq:p01}) and (\ref{eq:tE-1}) is the 
constant prefactor $\omega_0 = \sqrt{2/\pi} (v_T/a)$, with the thermal 
fluctuation velocity $v_T = \sqrt{2 E/M}$. 
Note that both $E_k$ and $E_e$ contribute to $E$ and thus to the collision
rate. Therefore, ${\cal C}_r$ can be split into a dynamic and a quasi-static 
contribution. 
The latter, the collision rate of elastic collisions is displayed in 
Fig.\ \ref{fig:tcn} in addition to the overall collision rate.
In an elastic collision, at least one collision partner had a collision
no longer than $t_c$ ago, see Eq.\ (\ref{eq:epsnew}). Therefore, smaller
$t_c$ values lead to lower collision rates (the decadic logarithm of 
$t_c = 10^{-3}$\,s, $10^{-4}$\,s, $10^{-5}$\,s, $10^{-6}$\,s, and $10^{-7}$\,s  
is given in the inset).

\subsection{Elastic particles, collisions, and energies}

In this subsection,
we are interested in $n_e=N_e/N$, the fraction of particles that are elastic,
i.e.~which had a collision no longer than $t_c$ ago. Furthermore,
we want to estimate $c_e={\cal C}_e/{\cal C}$, the fraction of collisions in 
which elastic particles participate, related to the elastic collision rate 
in the previous subsection. Finally, following the ideas in
subsection\ \ref{sec:energies}, we will split the total translational
energy into a kinetic and an elastic part, i.e.\ $E = E_k + E_e$
and determine $e_e=E_e/E$, the fraction of elastic energy in the system.

First we estimate the probability $p(t_c) = 1-N_e/N$ that a 
particle had {\em no} collision since a time $t_c$ \cite{reif88}. 
We know that $p(0)=1$, since no particle can suffer a collision
in zero time, and $p(\infty)=0$, since any particle will eventually 
collide (given that $a>0$ and that the system is not artificially frozen). 
Furthermore,
we also know the probability $t_E^{-1} {\rm d}t$ that a particle will 
collide within time ${\rm d}t$. Thus we have the probability
$1-t_E^{-1} {\rm d}t$ that the particle will {\em not} collide. Multiplying
the probability that it did not collide until $t_c$ with $1-t_E^{-1} {\rm d}t$
finally gives the probability that it does not collide until $t_c+{\rm d}t$.
First order Taylor expansion of 
$p(t_c+{\rm d}t) = p(t_c) (1 - t_E^{-1} {\rm d}t)$ 
around $t_c$ leads to $\frac{\rm d}{{\rm d}t}p(t) = p(t) t_E^{-1}$.
Integration for constant collision frequency $t_E^{-1}$ gives 
the fraction of `elastic' particles
\begin{equation}
n_e = 1 - p(t_c) = 1 - \exp( - t_c\,t_E^{-1})~.
\label{eq:ptc}
\end{equation}

\begin{figure}[hb]
\begin{center}
\epsfig{file=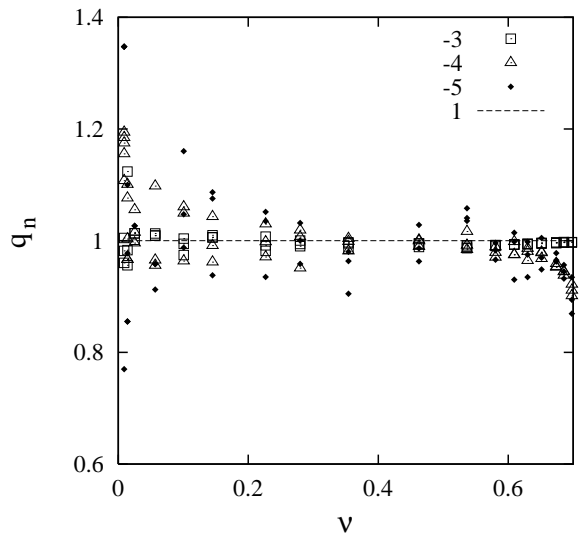,height=7.9cm,angle=-90,clip=}
\end{center}
\caption{
The quality factor $q_n$, i.e.~the ratio of $\left < N_e \right > / N$
and $n_e$ from Eq.\ (\ref{eq:ptc}), plotted against $\nu$. 
The simulations are the same as in
Fig.\ \ref{fig:etap}. Values close to unity indicate good agreement
between theory and simulations for the $t_c$ values used. To distinguish
between different $t_c$ values, $\log_{10} t_c$ is given in inset.
}
\label{fig:ne}
\end{figure}

In Fig.\ \ref{fig:ne} we present the quality factor 
$q_n = \left < N_e \right > / (N n_e)$, the ratio of the measured 
fraction of elastic particles and of the analytical expression
from Eq.\ (\ref{eq:ptc}). 
Each point corresponds to an average over 30
snapshots from simulations with $N=1435$ particles. 

Data for different 
$t_c$ values can be obtained from the same simulation since each particle 
`remembers' when it had its last collision at time $t_n^{(i)}$ and
because no dissipation is active. Dissipation would make the sequence of 
collisions dependent on $t_c$.
The agreement between simulations and the theoretical
prediction, Eq.\ (\ref{eq:ptc}), is reasonable ($q_n \approx 1$). 
Deviations occur for small volume fractions
$\nu$ due to bad statistics and for large volume fractions $\nu > 0.7$ due 
to the transition to the ordered system. 

\begin{figure}[ht]
\begin{center}
\epsfig{file=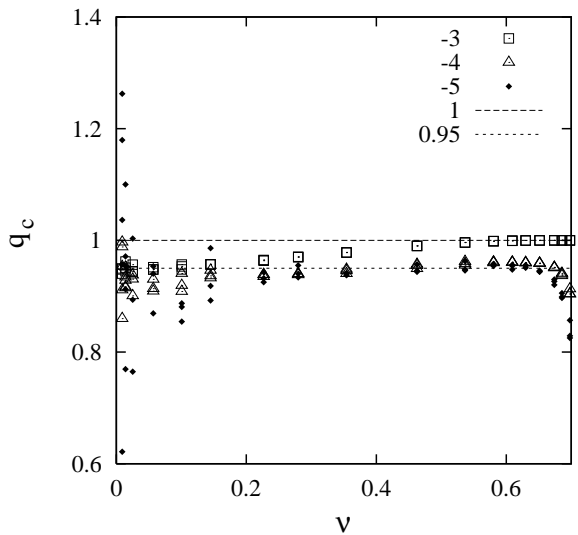,height=7.9cm,angle=-90,clip=}
\end{center}
\caption{
The ratio of simulational and theoretical results on the probability
for elastic collisions plotted against $\nu$. Exact agreement would
correspond to $q_c = 1$.
}
\label{fig:tn}
\end{figure}

However, Eq.\ (\ref{eq:ptc}) is {\em not} the fraction of elastic collisions 
$c_e$ which instead has to be computed as the sum of probabilities that either 
both or only one of the collision partners belongs to the elastic particles.
Thus one gets 
\begin{equation}
c_e = n_e^2 + 2 n_e ( 1 - n_e ) = 1 - \exp( - 2 t_c\,t_E^{-1})~,
\label{eq:Ce}
\end{equation}
almost in agreement with the simulations as displayed in Fig.\ \ref{fig:tn}.
The quality factor 
$q_c = \left < {\cal C}_e \right > / (\left < {\cal C} \right > c_e)$
plotted against $\nu$ indicates a difference between the measured and
the calculated fraction of elastic collisions.
Interestingly, the simulation data show that there occur about five 
percent less elastic collisions than expected from Eq.\ (\ref{eq:Ce}).
The data presented in Fig.\ \ref{fig:tn} were obtained by averaging
$\left < {\cal C}_e \right > / \left < {\cal C} \right >$ over a time 
interval $\Delta t$ large enough to allow for more than about $10^4$
collisions per particle. A more rigorus theoretical treatment that would
lead to the exact value of $c_e$ is beyond the scope of this study
and will be discussed elsewhere \cite{luding98g}. 

\begin{figure}[ht]
\begin{center}
\epsfig{file=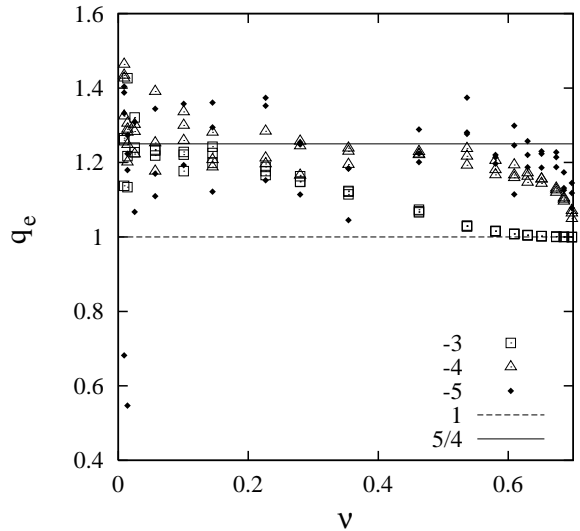,height=7.9cm,angle=-90,clip=}
\end{center}
\caption{
The ratio of simulational and theoretical results on the fraction of
elastic energy in the system plotted against $\nu$. A quality factor 
$q_e = 1$ would correspond to exact agreement. The value $5/4$ was
taken from a more elaborate kinetic theory calculation \protect\cite{luding98g}.
}
\label{fig:tnE}
\end{figure}

Finally, we estimate the elastic energy contained in the system in a 
simple minded, mean-field way. The fraction of elastic energy $e_e$ should
be the product of $n_e$ and $E$ when all particles would have the same
(mean) energy and would contribute to $E_e$ with the same probability.
Unfortunately, the simulation results in Fig.\ \ref{fig:tnE} indicate
that the mean-field approach is not valid. The discrepancy between 
the mean-field estimate presented above for $n_e$ and the
numerical simulations can, however, be understood via qualitative arguments.
Particles with greater velocities have a higher probability to collide
and, therefore, have a greater collision rate than slower particles.
Due to the greater collision rate, fast particles are more
likely to contribute to $E_e$ and $E_e$ is increased since faster
particles contribute with a greater energy \cite{luding98g}.

\section{The TC model in dissipative systems}

\label{sec:cluster}

One of the phenomena which received a great deal of attention
in the last years is the clustering instability in rather dilute
systems of dissipative particles
\cite{goldhirsch93,mcnamara96,luding98b}.
Initially given a homogeneous
density and a Maxwellian ve\-lo\-city-distribution, the system cools
due to dissipation \cite{mcnamara96,luding98d}.
This cooling regime is unstable to
perturbations with large enough wavelength, i.e.~small enough
wavenumber \cite{mcnamara96,brey96}. The homogeneous state can
be well described by hydrodynamic theory, but the standard description
breaks down as soon as the perturbations grow -- assumptions like 
homogeneity or ``molecular chaos'' are not longer true 
\cite{luding98b,luding98e}. Furthermore, it has been 
recognized that a hydrodynamic description breaks down 
due to the divergence of the collision rate and the connected 
dramatic decrease of free volume during the inelastic collapse
\cite{du95,grossman97}.\\
The instability may be understood in a qualitative manner:
The homogeneous state contains thermal velocity fluctuations.  A convergent
velocity fluctuation leads to increasing densities in certain regions.
As the collision rate increases in these regions, so does the energy
dissipation rate. If the energy dissipation rate is great enough, 
the pressure cannot reverse the convergent flow. If this process is 
not terminated by sufficiently strong perturbations, it is 
self-stabilizing, causes clusters, and may eventually lead to the
inelastic collapse.\\
The clustering instability and the inelastic collapse were carefully 
examined in 1D
\cite{bernu90,mcnamara92,mcnamara93,luding94,grossman96b,kudrolli97,kudrolli97b}
and in 2D 
\cite{sibuya90,goldhirsch93,mcnamara94,trizac95,mcnamara96,spahn97,deltour97,orza97}.
Cluster growth could be described theoretically in the case of irreversible 
aggregation ($r=0$) \cite{trizac95,trizac96} and, more recently, also a 
theory for the growth of density fluctuations was proposed for $r \ne 0$
\cite{brito98}.
Detailed examination of the inelastic collapse by McNamara and Young
\cite{mcnamara96} led to the picture of different `phases'. In a periodic 
system without external forcing exists a critical dissipation -- connected to 
system size, volume fraction and restitution coefficient -- above which 
clustering occurs and below which the system stays in molecular chaos.
In the transient regime shearing modes or large scale eddies are frequently 
observed. The case of homogeneous cooling is rather well understood
\cite{luding98e} so that we mainly focus on situations with rather strong
dissipation when the system is no longer homogeneous.

Using event-driven simulations, periodic 2D systems with side-length 
$L = l/(2a)$ in 2D are examined in the following. A system
contains $N$ particles of radius $a$ and volume fraction 
$\nu = N \pi (a / l)^2 = (\pi/4) N/L^2$. 
In this section the particles are dissipative with a restitution 
coefficient $r$. We apply the TC model as described above, 
i.e.~we use Eq.\ (\ref{eq:epsnew}) with
the parameter $t_c$ to be specified. Initially we arrange the particles on an
ordered lattice and give each of them a random velocity; then the system is 
equilibrated with $r=1$ until a Maxwellian velocity distribution is obtained.
Finally, dissipation is switched on and the simulation starts at $t=0$\,s. 

\subsection{Inhomogeneous cooling -- Finite size effects}

\begin{figure}[htb]
\epsfig{file=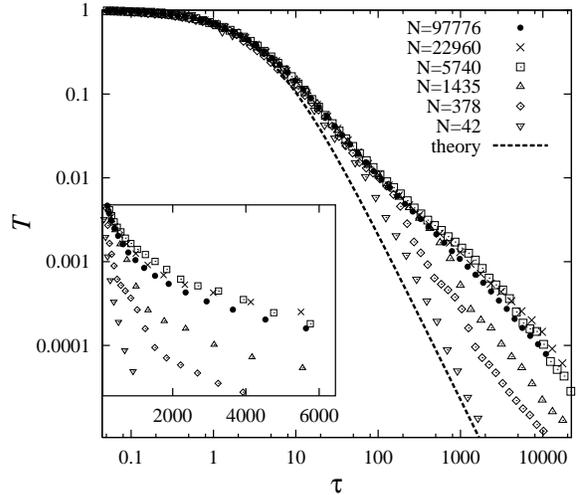,width=7.0cm,angle=-90}
\caption{
$T$ as function of $\tau$ for simulations with $N=97776$, 22960, 5740, 1435, 
378, and 42, $\nu = 0.227$, $r=0.4$, and $t_c = 10^{-5}$\,s. The dashed line 
gives the solution of the homogeneous cooling state 
Eq.\ (\protect\ref{eq:HCS}). 
}
\label{fig:cluster_Ka}
\end{figure}

In this subsection we discuss typical simulations with volume fraction
$\nu = 0.227$, restitution coefficient $r=0.40$, and contact duration 
$t_c = 10^{-5}$\,s. The system size is varied so that $N=97776$, 22960, 
5740, 1435, 378, and 42 particles fit into the system. 
In Fig.\ \ref{fig:cluster_Ka},  
the dimensionless kinetic energy $T=E(t)/E(0)$ is plotted against
the rescaled time $\tau=t_E^{-1}(0)\,t$, with the initial collision 
rate $t_E^{-1}(0) \approx 444$\,s$^{-1}$ for all simulations. 
\begin{figure}[htb]
\epsfig{file=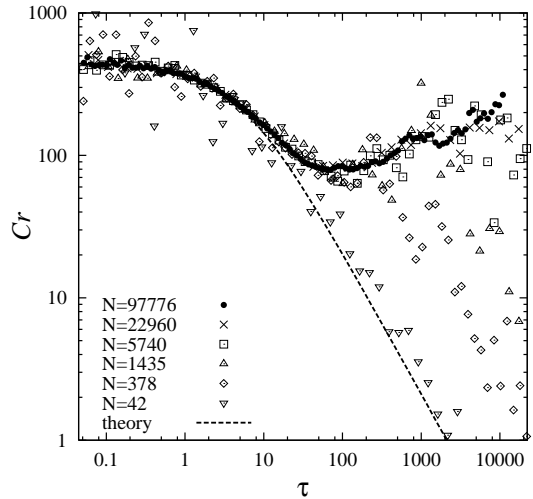,width=7.0cm,angle=-90}
\caption{
Collision rate ${\cal C}_r$, see Eq.\ (\protect\ref{eq:Cn}), as function
of $\tau$ from the same simulations as in Fig.\ \protect\ref{fig:cluster_Ka}.
The dashed line gives the solution of the homogeneous cooling state.
}
\label{fig:cluster_Kb}
\end{figure}

Following the estimate of a critical restitution coefficient,
see Eq.\ (9) in Ref.\ \cite{mcnamara96}, below which the inelastic
collapse can be expected, we compute for our simulations
\begin{equation}
r_c(N) \approx \tan ^2 \left [ \frac{\pi}{4}
            \left ( 1-\frac{1}{\lambda_{\rm opt}} \right ) \right ]~,
\end{equation}
with the non-dimensional optical depth 
$\lambda_{\rm opt} = \sqrt{\pi N \nu}/2$.
Comparing the critical restitution coefficients $r_c(97776)=0.976$, 
$r_c(22960)=0.952$, $r_c(5740)=0.906$, $r_c(1435)=0.821$, $r_c(378)=0.680$ 
and $r_c(42)=0.296$ to $r=0.40$ used for the simulations, we note that
all are larger, $r_c > r$, except for the tiny system with $N=42$. 
However, due to the TC model used in our simulation no collapse is
observed.
The dashed line in Fig.\ \ref{fig:cluster_Ka} gives the analytical 
solution for the homogeneous cooling state (HCS) of smooth particles 
(see \cite{luding98d} and references therein):
\begin{equation}
T = \left ( {1+\frac{1-r^2}{4} ~ \tau} \right )^{-2}~.
\label{eq:HCS}
\end{equation}
Only the smallest system is close to the theoretical prediction,
all others deviate stronger with increasing system size.

The collision rate in the homogeneous cooling state is
directly linked to the fluctuation velocity $v_T$, see
Eq.\ \ref{eq:tE-1}. Because $v_T$ is the only quantity 
that changes during the simulations, we can also express
the collision rate in terms of $\sqrt{T}=v_T/v_0$ so that
due to normalization $T_0=1$, one has 
${\cal C}_r = {\cal C}_r(0) \sqrt{T}$. In order to test 
this prediction, we plot ${\cal C}_r$ against $\tau$
in Fig.\ \ref{fig:cluster_Kb} from the same simulations
as in Fig.\ \ref{fig:cluster_Ka}.

As can be seen in both figures \ref{fig:cluster_Ka} and \ref{fig:cluster_Kb},
for small $\tau < 5$ the simulations agree with the theory. 
For larger $\tau$ deviations from the homogeneous cooling regime
occur and the energy decay slows down due to the density instability
and the build-up of clusters. The behavior of the energy as a 
function of time is independent of the system size up to 
$\tau \approx 200$ when the small system $N=378$ starts to deviate
from the larger systems. The deviation from the common behavior 
occurs later with increasing system size, indicating finite-size 
effects. A closer examination of snapshots from the simulations 
leads to the conclusion that the deviation from the slow cooling
regime occurs when the clusters have reached the system size.
The tiny system $N=42$ is anyway too small to allow large scale
structures and therefore follows the HCS prediction more closely.
\begin{figure*}[htb]
{\noindent $\tau=89$ \hfill $\tau=1422$ \hfill $\tau=23110$ \vspace{.1cm}}\\
\hspace{ 0.0cm} \epsfig{file=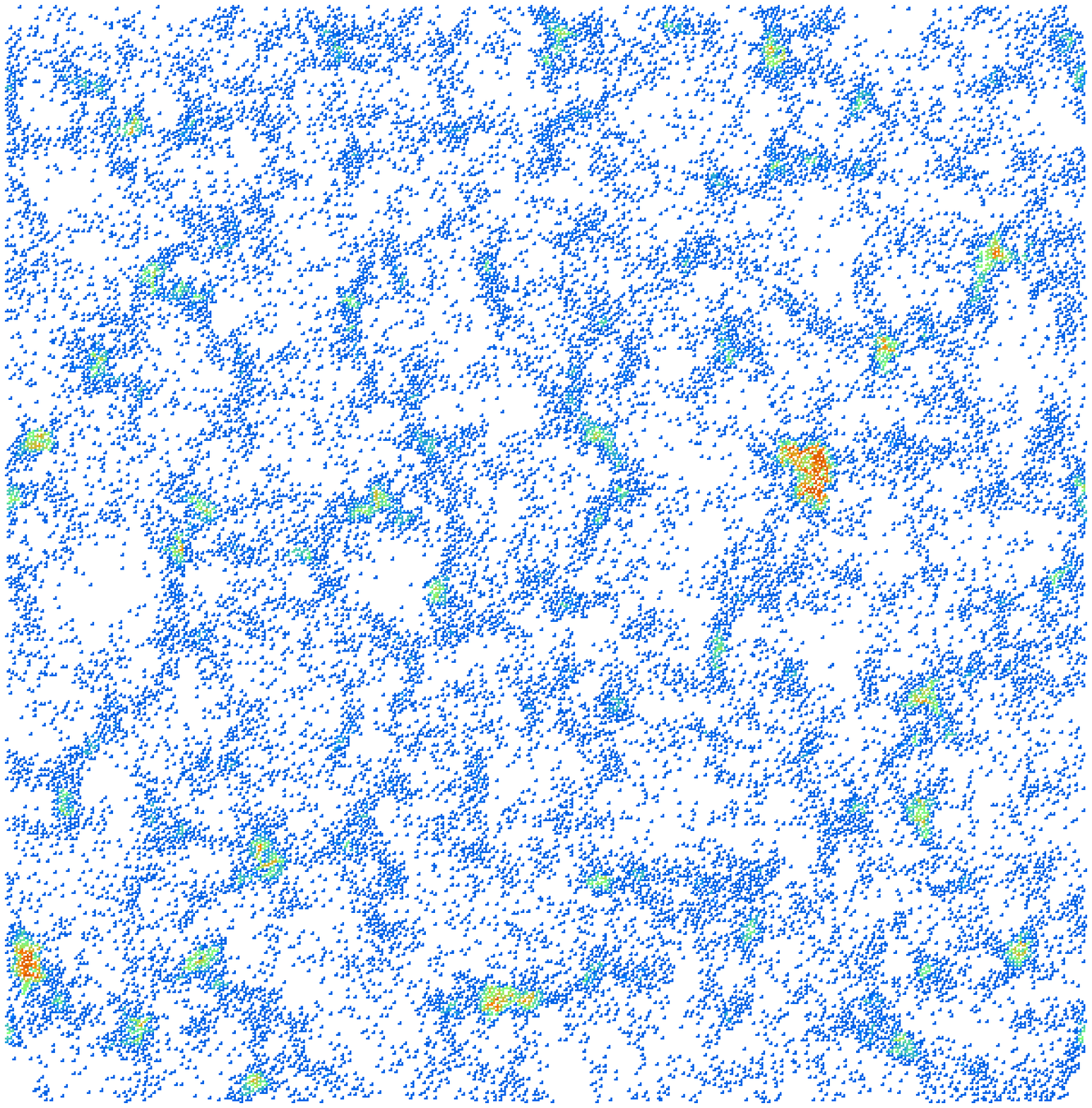,height=6.0cm} \hfill
\hspace{-0.2cm} \epsfig{file=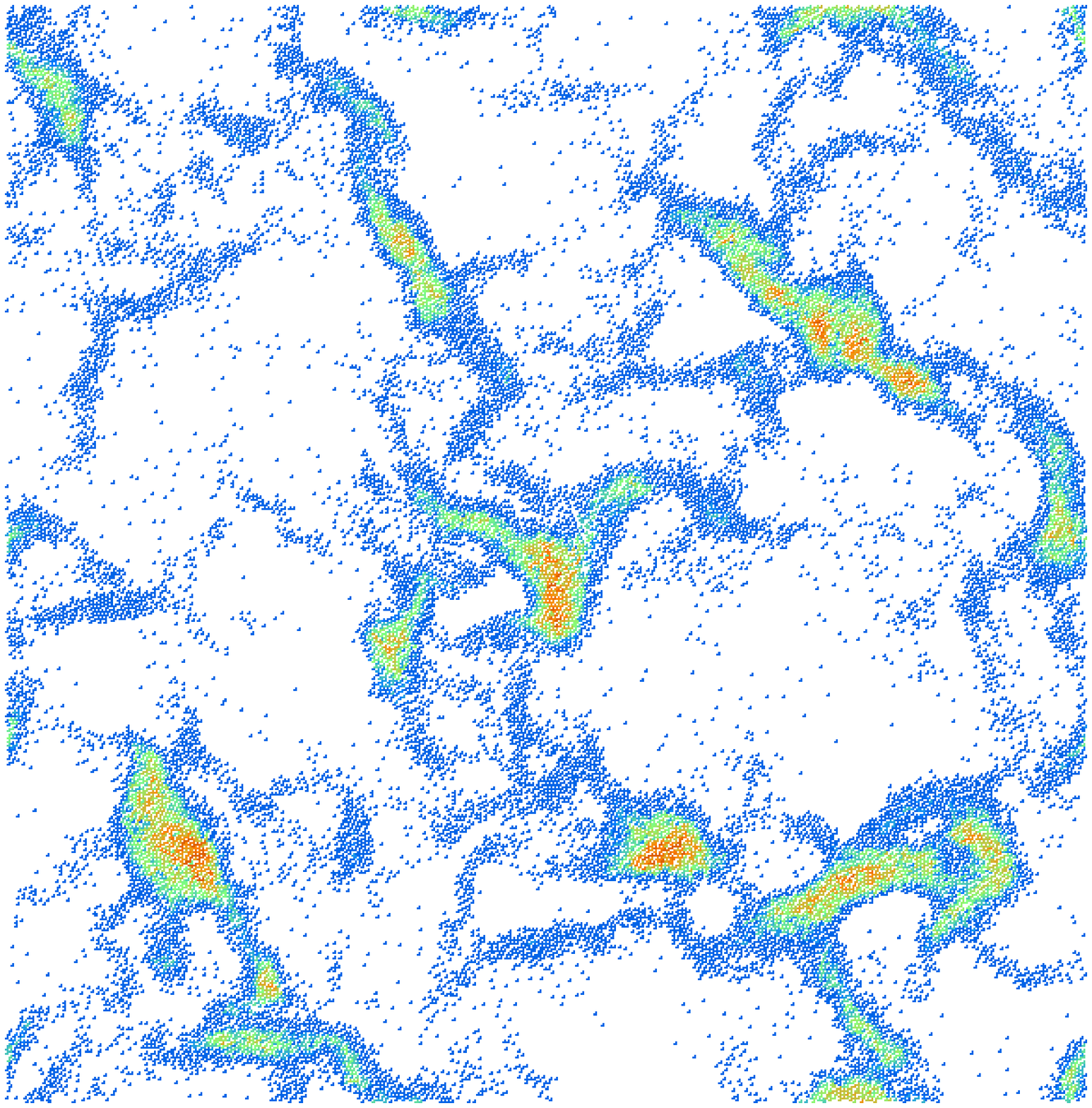,height=6.0cm} \hfill
\hspace{-0.2cm} \epsfig{file=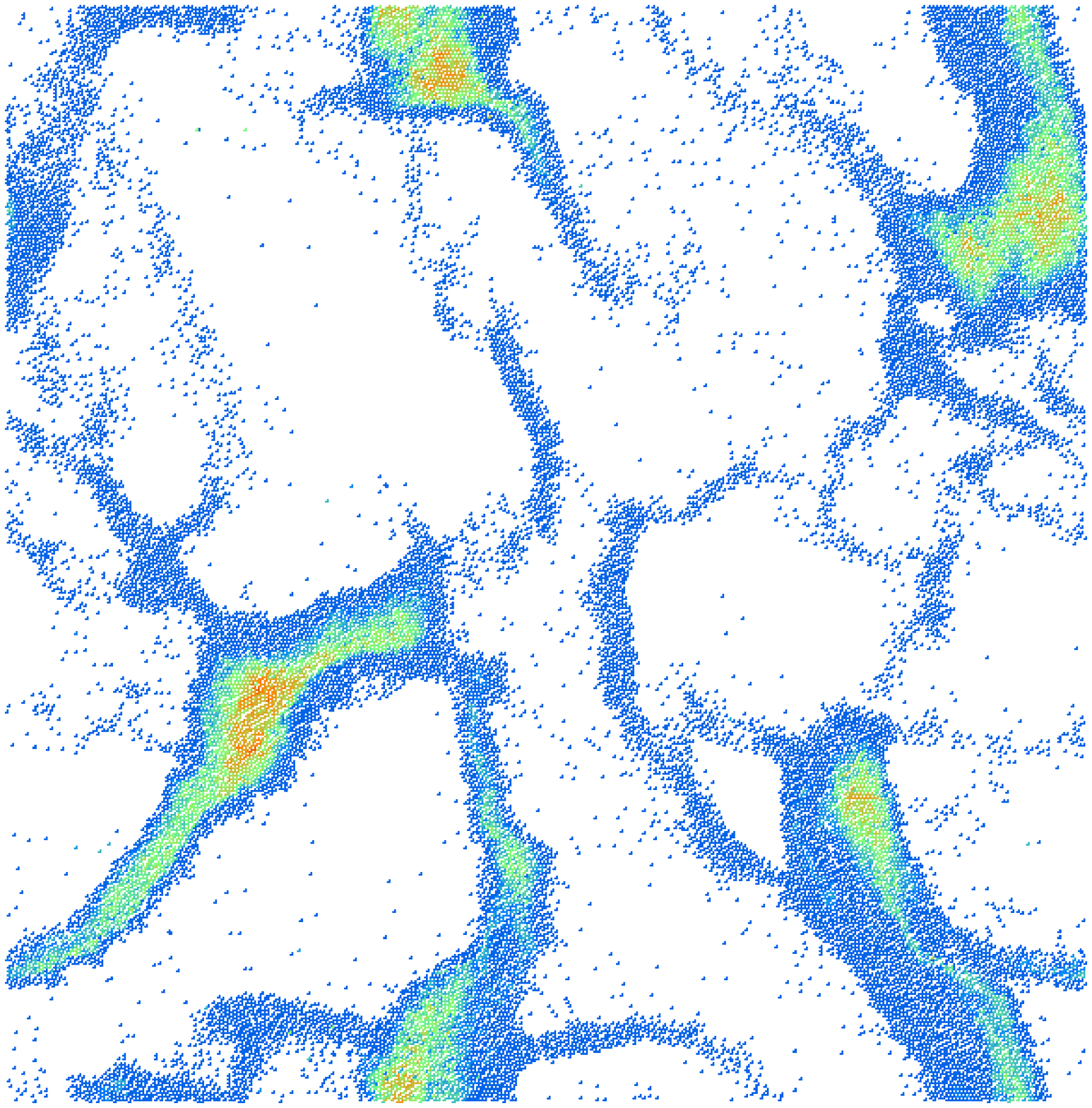,height=6.0cm}\\
\caption{ED simulation with $N=22960$ particles in a system of
size $L=280$, volume fraction $\nu = 0.227$, restitution coefficient
$r=0.4$, and contact duration $t_c = 10^{-5}$\,s.
The collision rate is color-coded red (${\cal C} > 1100$\,s$^{-1}$), green
(${\cal C} \approx 700$\,s$^{-1}$), and blue (${\cal C} < 300$\,s$^{-1}$).
}
\label{fig:cluster_nc}
\end{figure*}
\begin{figure*}[hbt]
{\noindent $t_c=10^{-2}$\,s} \hfill {$t_c=10^{-3}$\,s} 
\hfill {$t_c=10^{-4}$\,s} 
\hfill {$t_c=10^{-5}$\,s} 
\vspace{.1cm} \\
\epsfig{file=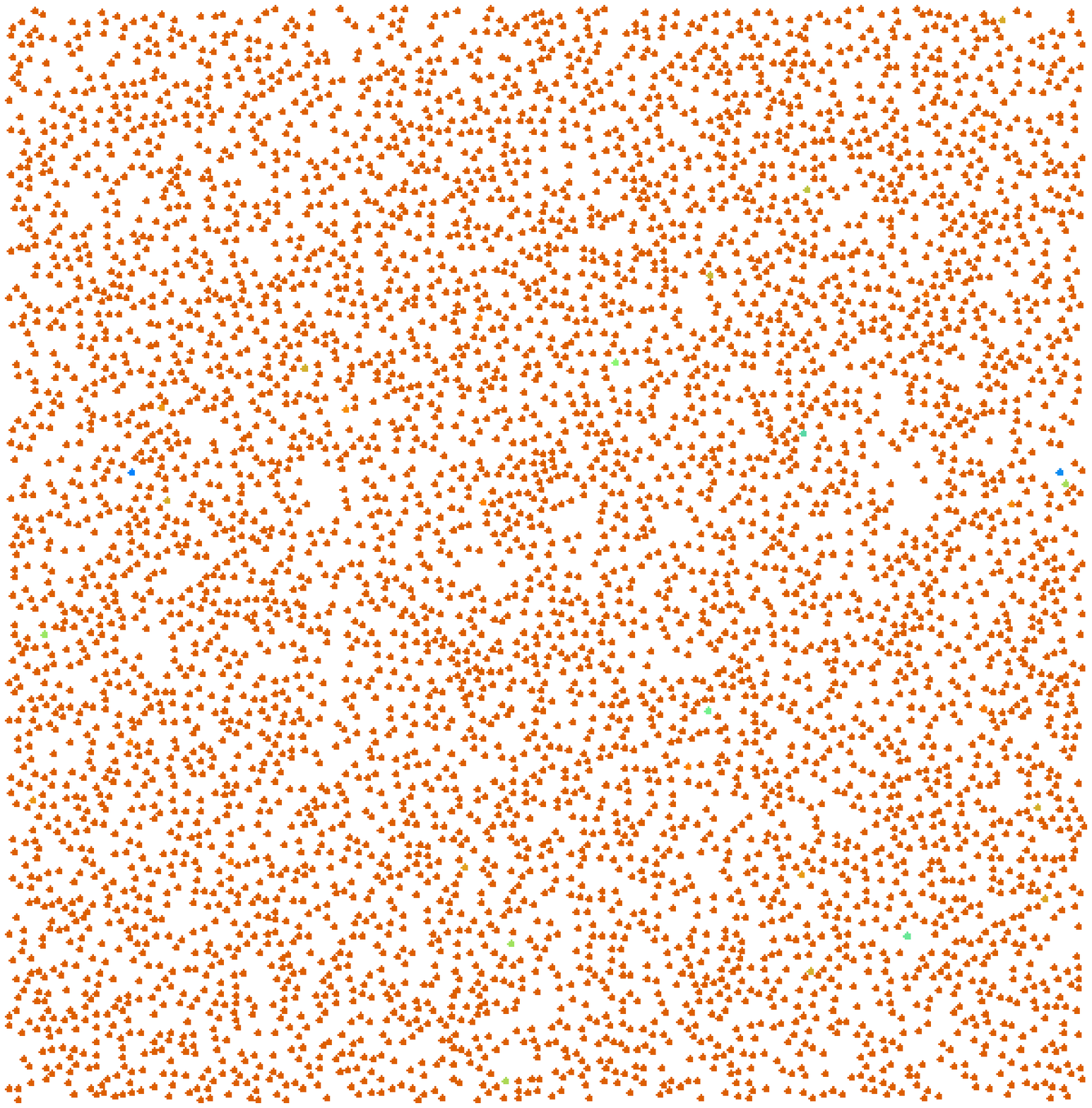,height=4.5cm} \hfill
\hspace{-0.2cm} \epsfig{file=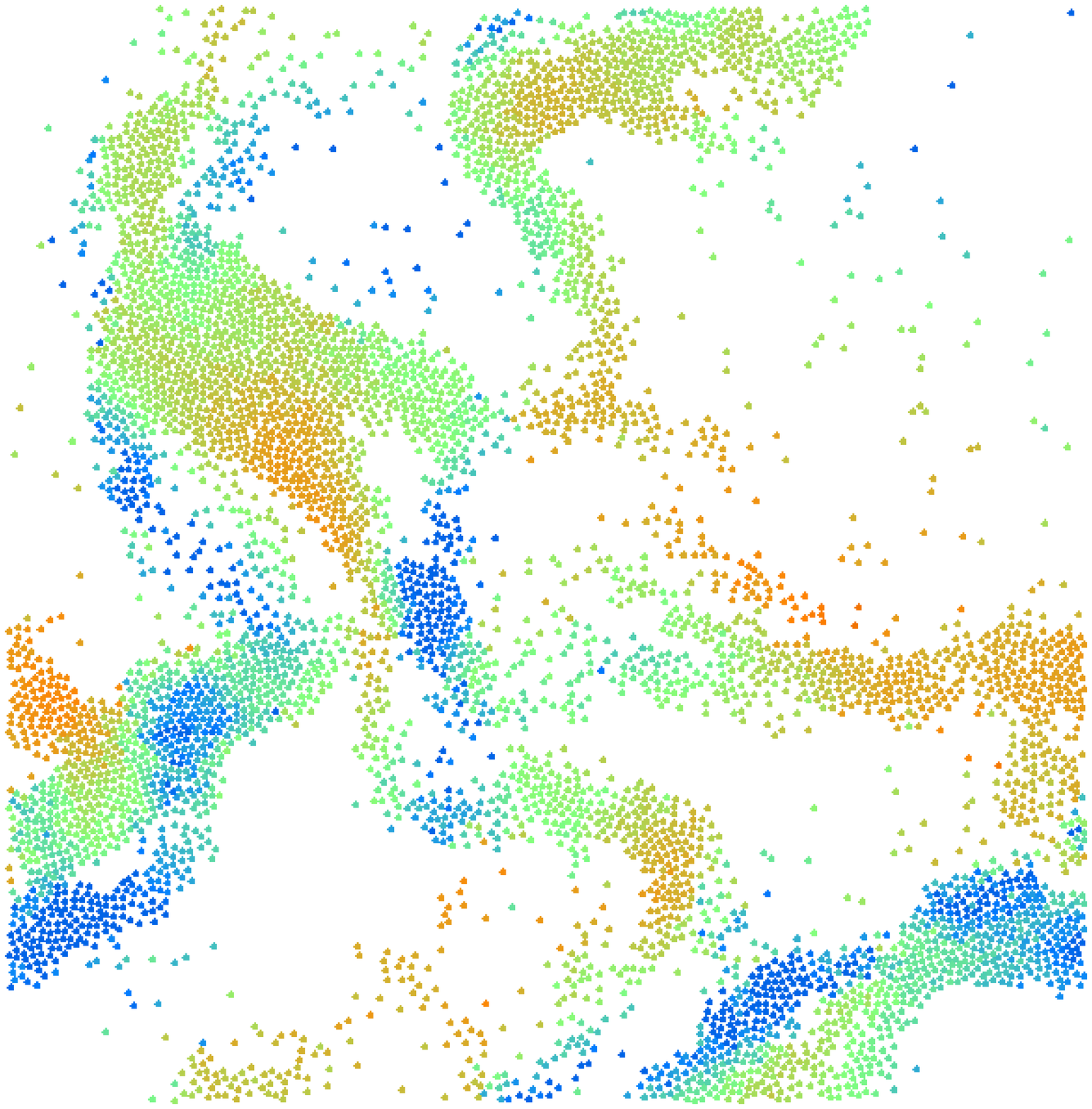,height=4.5cm} \hfill
\hspace{-0.2cm} \epsfig{file=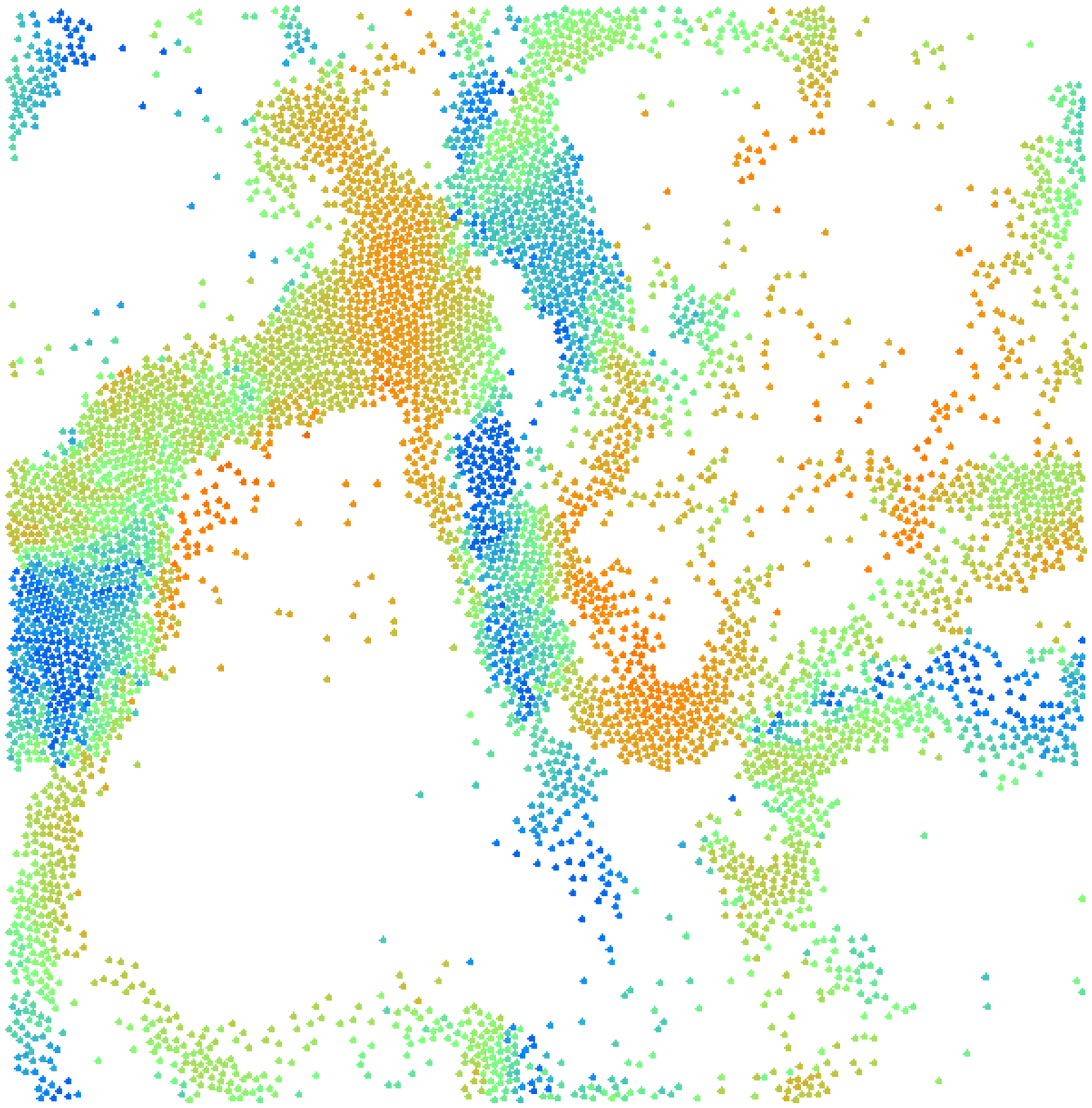,height=4.5cm} \hfill
\hspace{-0.2cm} \epsfig{file=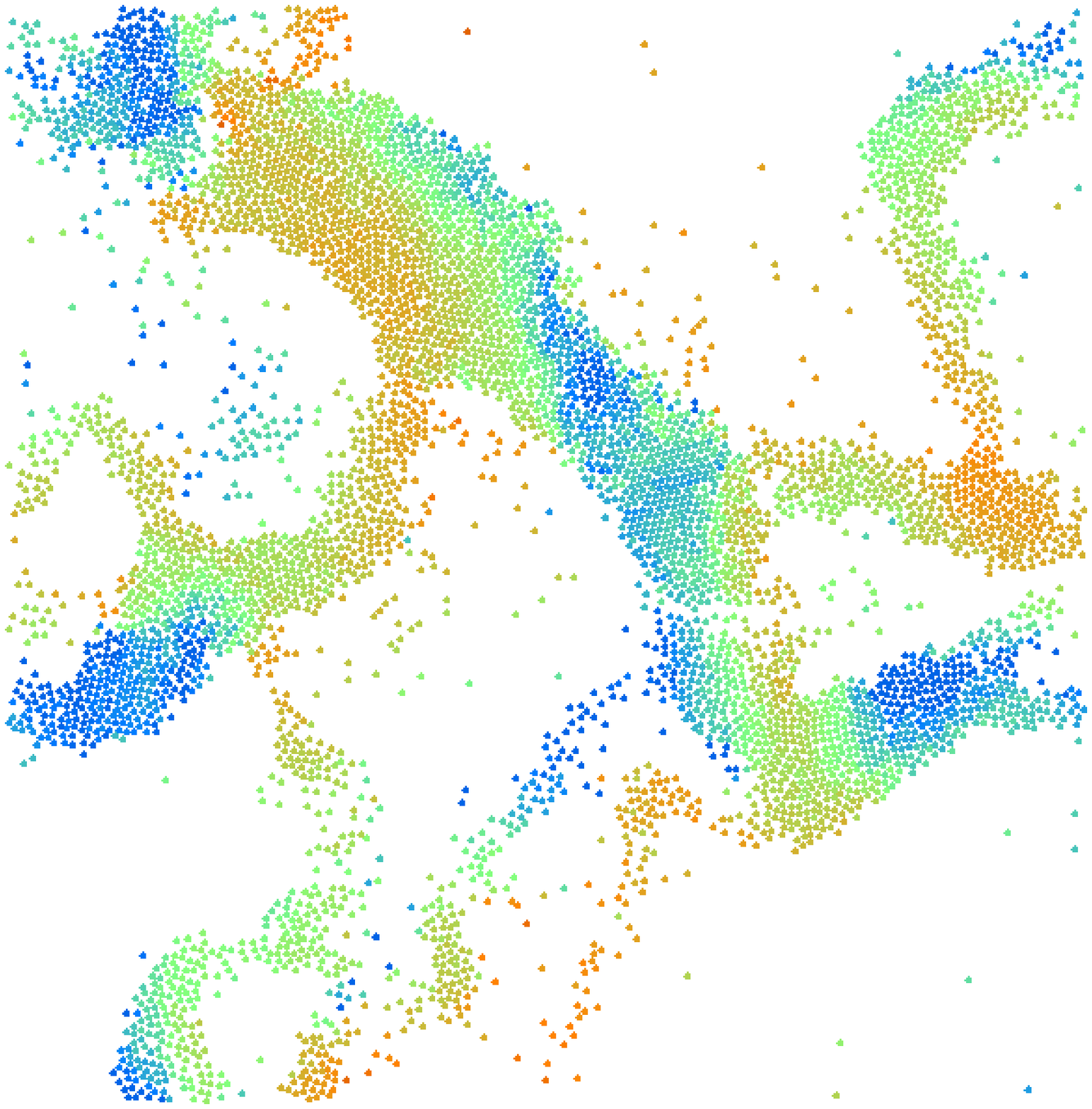,height=4.5cm} \\
{$t_c=10^{-6}$\,s} \hfill 
{$t_c=10^{-8}$\,s} \hfill 
{$t_c=10^{-10}$\,s} \hfill {$t_c=10^{-12}$\,s \vspace{.1cm}} \\
\epsfig{file=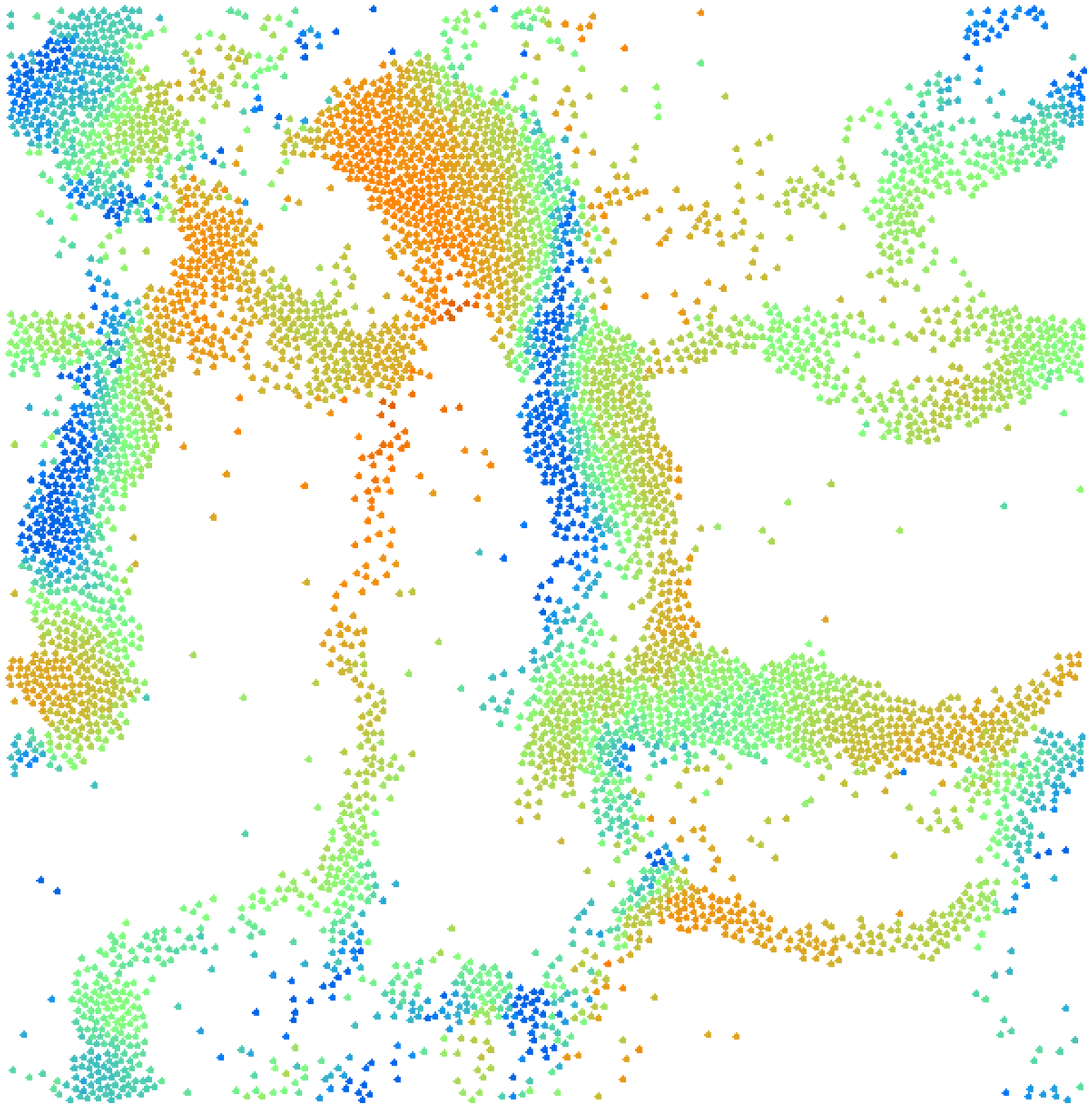,height=4.5cm} \hfill
\hspace{-0.2cm} \epsfig{file=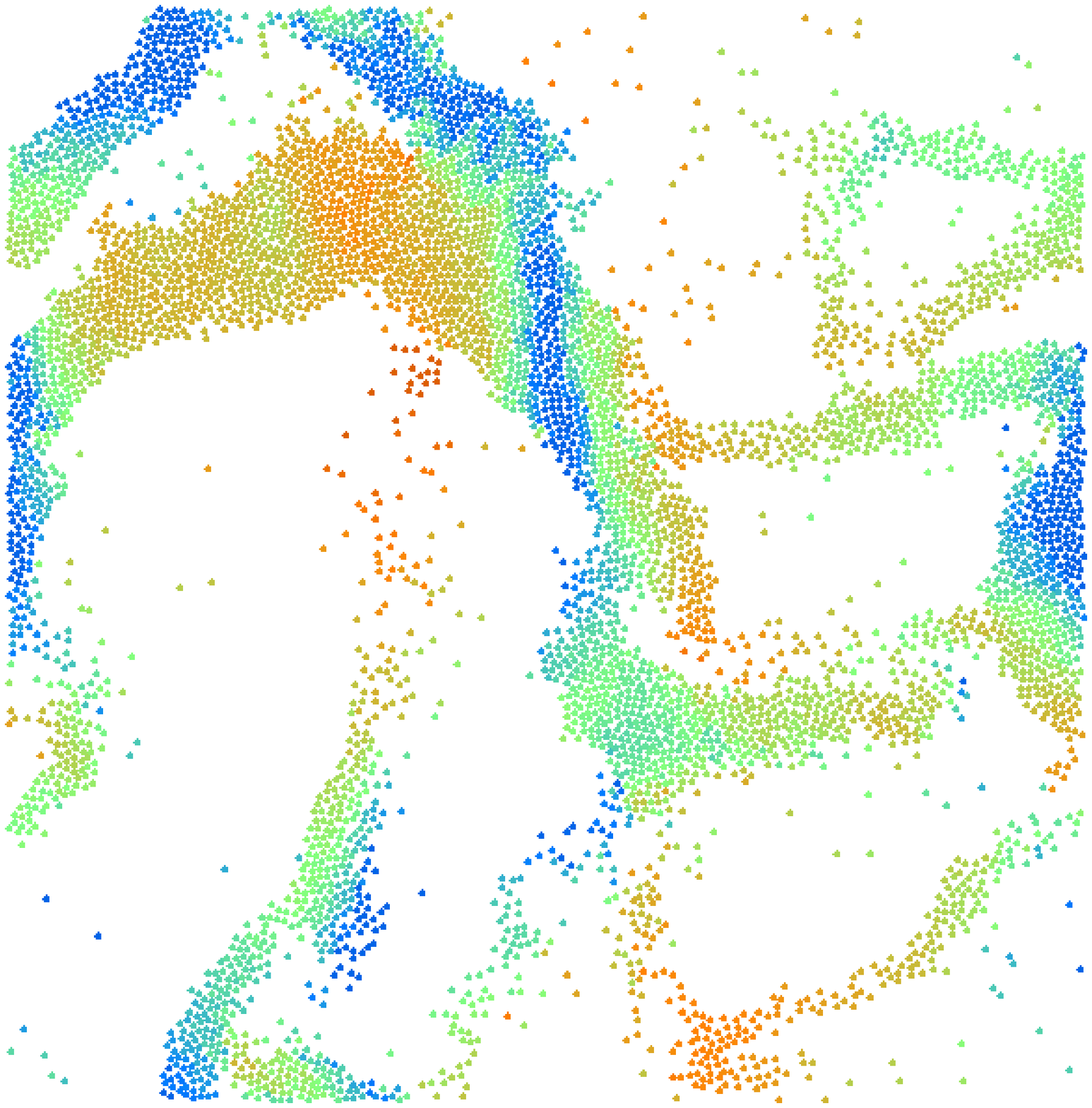,height=4.5cm} \hfill
\hspace{-0.2cm} \epsfig{file=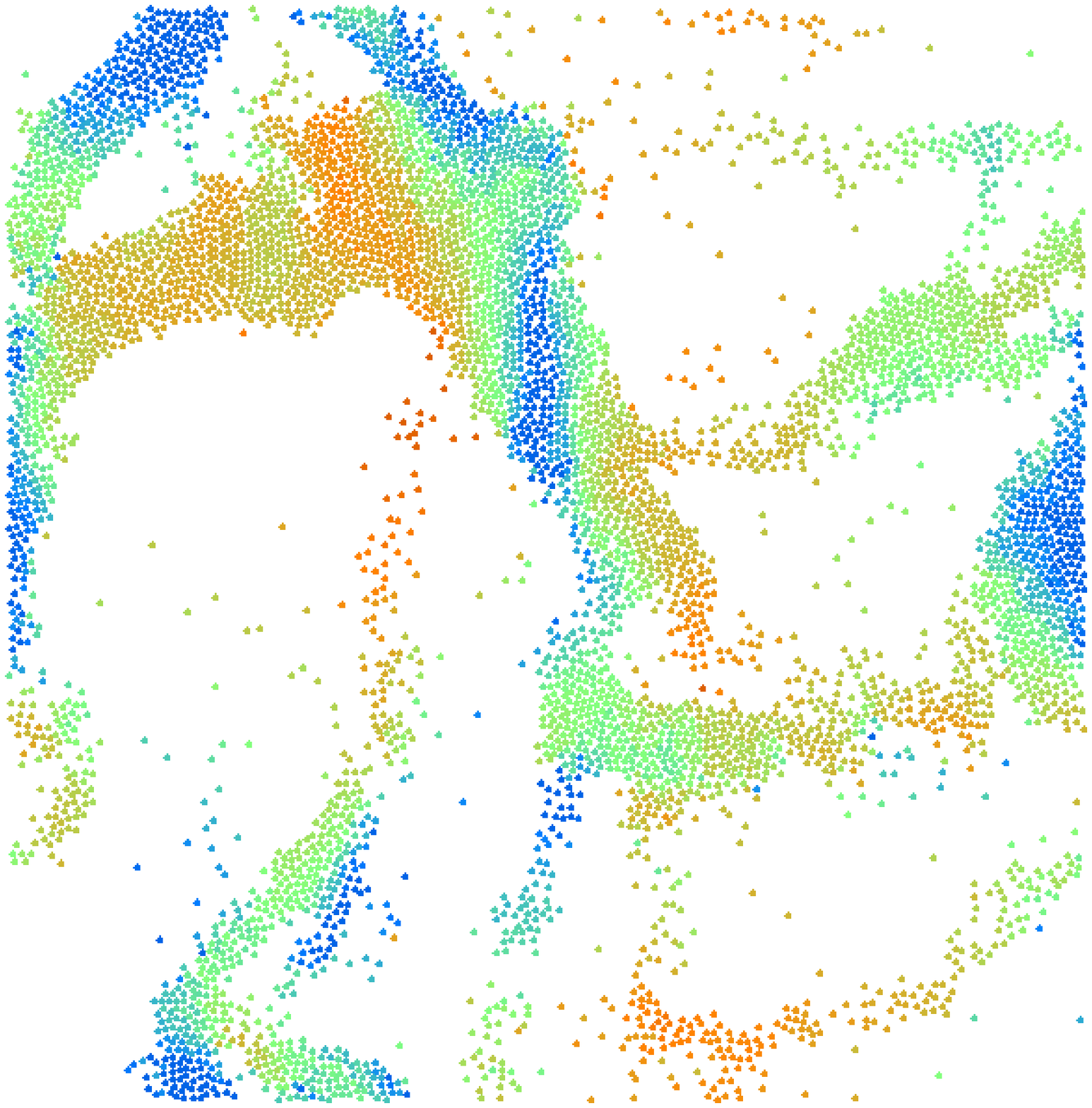,height=4.5cm} \hfill
\hspace{-0.2cm} \epsfig{file=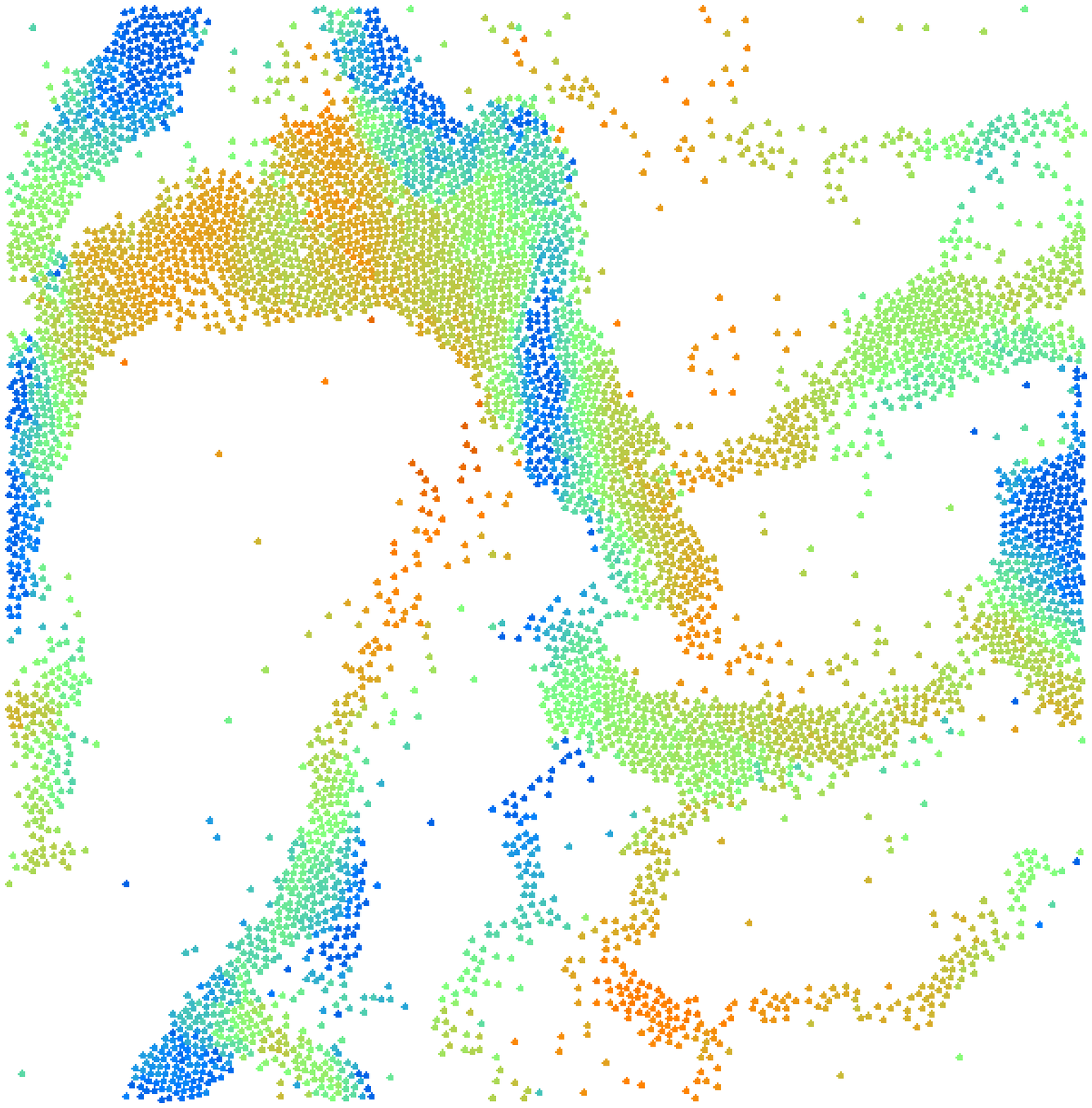,height=4.5cm} \\
\caption{Snapshots at time $t=5$\,s, i.e.~$\tau=2222$, from
ED simulations started from identical initial
conditions, with $N=5740$, $L=140$, $\nu = 0.227$, $r=0.4$, and different contact duration
$t_c$. The mean fluctuation velocity (relative to the center of mass) is color-coded
red ($v_T/v_0 > 0.045$), green ($v_T/v_0 \approx 0.02$), and blue ($v_T/v_0 < 0.01$),
with initial mean velocity $v_0 = v_T(0) = 0.3615$\,m\,s$^{-1}$.
}
\label{fig:cluster_e}
\end{figure*}

In Fig.\ \ref{fig:cluster_nc} snapshots of the simulation with $N=22960$ from 
Fig.\ \ref{fig:cluster_Ka} are displayed. With increasing time $\tau$  
structures build up in the system and grow in size. In the red and
green regions in the centers of the clusters the collision rate is 
largest. The size of the clusters grows with time and the qualitative
cooling behavior of the system changes as soon as the clusters get to 
be as large as the system.

\subsection{Variation of the contact duration}

In the following set of simulations, $t_c$ is varied while the other
parameters are fixed to $N=5740$, $\nu \approx 0.227$, and $r=0.4$.
In Fig.\ \ref{fig:cluster_e} snapshots from the simulations with
different contact duration are displayed. The simulations were 
set up with identical initial conditions and the snapshots were 
taken at the same time. For $t_c=10^{-2}$\,s cooling is almost
hindered, since the time between collisions is much greater than
the dissipation cut-off $t_c$. The simulations with $t_c \le 10^{-3}$\,s
all show density instabilities and clusters, however, the pictures
differ in details. The change in behavior between $t_c=10^{-2}$\,s and
$10^{-3}$\,s will be dicussed in more detail elsewhere \cite{luding98g}. 
Only for very small $t_c \le 10^{-8}$\,s the 
simulations lead to almost identical results.
The color of the snapshots represents the 
particle velocities in the center of mass reference frame. A large cluster
with different colors is thus not a block without internal motion
but rather a liquid-like arrangement with strong internal shearing.

\begin{figure}[htb]
\epsfig{file=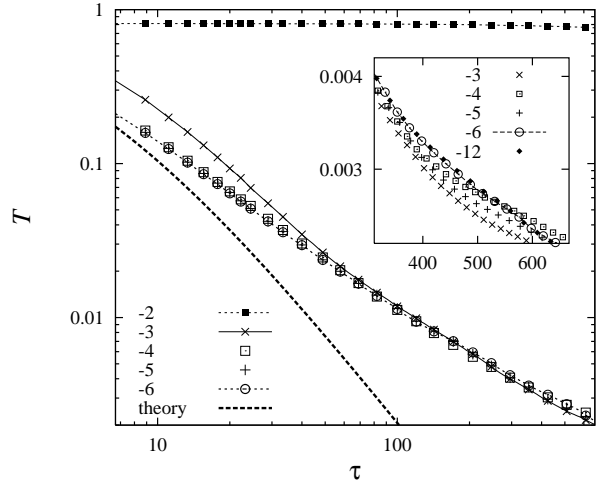,height=8.0cm,angle=-90} 
\caption{
Dimensionless energy $T$ plotted against $\tau$ from simulations
with $r = 0.4$ and $\log_{10} t_c$ as given in the plot. The inset is
a zoom into the plot at large $\tau$ and small $T$. The dashed
line corresponds to Eq.\ (\protect\ref{eq:HCS}) with $r=0.4$.
}
\label{fig:tcfig01a}
\end{figure}

In Fig.\ \ref{fig:tcfig01a}, the dimensionless kinetic energy 
$T$ is plotted against the rescaled time $\tau=t_E^{-1}(0)\,t$,
with the initial collision rate $t_E^{-1}(0) \approx 444$\,s$^{-1}$.
As contact duration, various values between $t_c = 10^{-2}$\,s,
and $10^{-12}$\,s were used ($\log_{10} t_c$ is given in the inset
for identification). The simulations with $t_c < 5 \times 10^{-4}$\,s
lead to the same functional behavior of $T$ as function of $\tau$,
except for small deviations for large $\tau$, as can be obtained
from the inset. However, the data with $t_c \le 10^{-6}$\,s cannot
be distinguished, i.e.~for small enough $t_c$ the energy of the
system is not influenced by the contact duration. For very small
$\tau < 5$, the simulations agree with the theoretical result
for the homogeneous cooling state from Eq.\ (\ref{eq:HCS}).
For larger $\tau$ deviations from the homogeneous cooling regime
occur and the energy decay slows down due to the density instability
and the build-up of clusters.

\begin{figure}[htb]
\epsfig{file=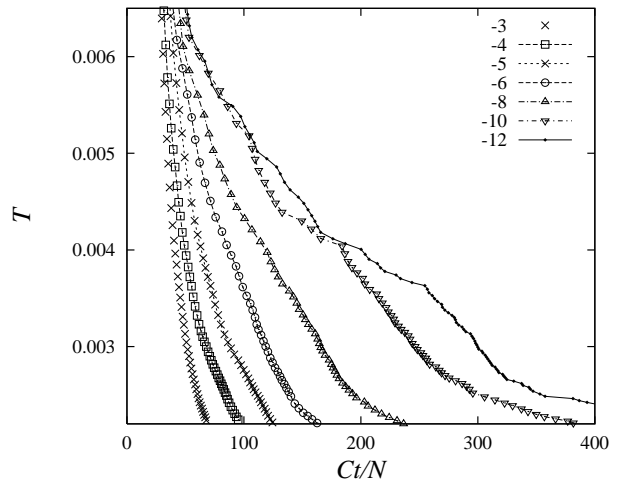,height=8.0cm,angle=-90}
\caption{
$T$ plotted against $C_t/N$ for the same simulations as in 
Fig.\ \protect\ref{fig:tcfig01a}.
}
\label{fig:tcfig01b}
\end{figure}

\begin{figure*}[htb]
~~~~~ {\large (a)} \hfill~\hfill
     {\large (b)} \hfill~\hfill 
     {\large (c)} ~~~~~~~~~~~ \hfill ~ \\
\epsfig{file=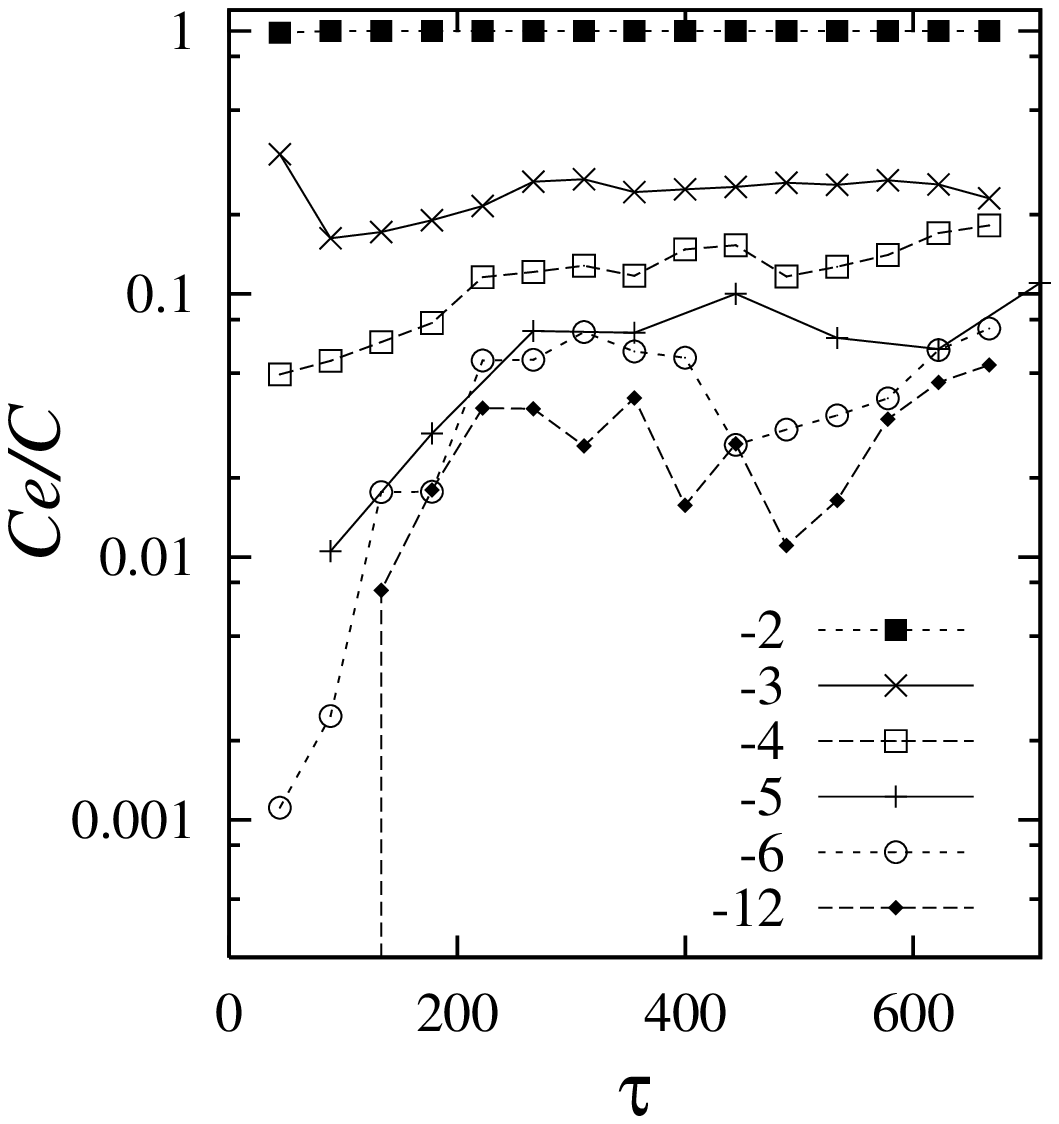,height=5.8cm,angle=-90} \hfill
\epsfig{file=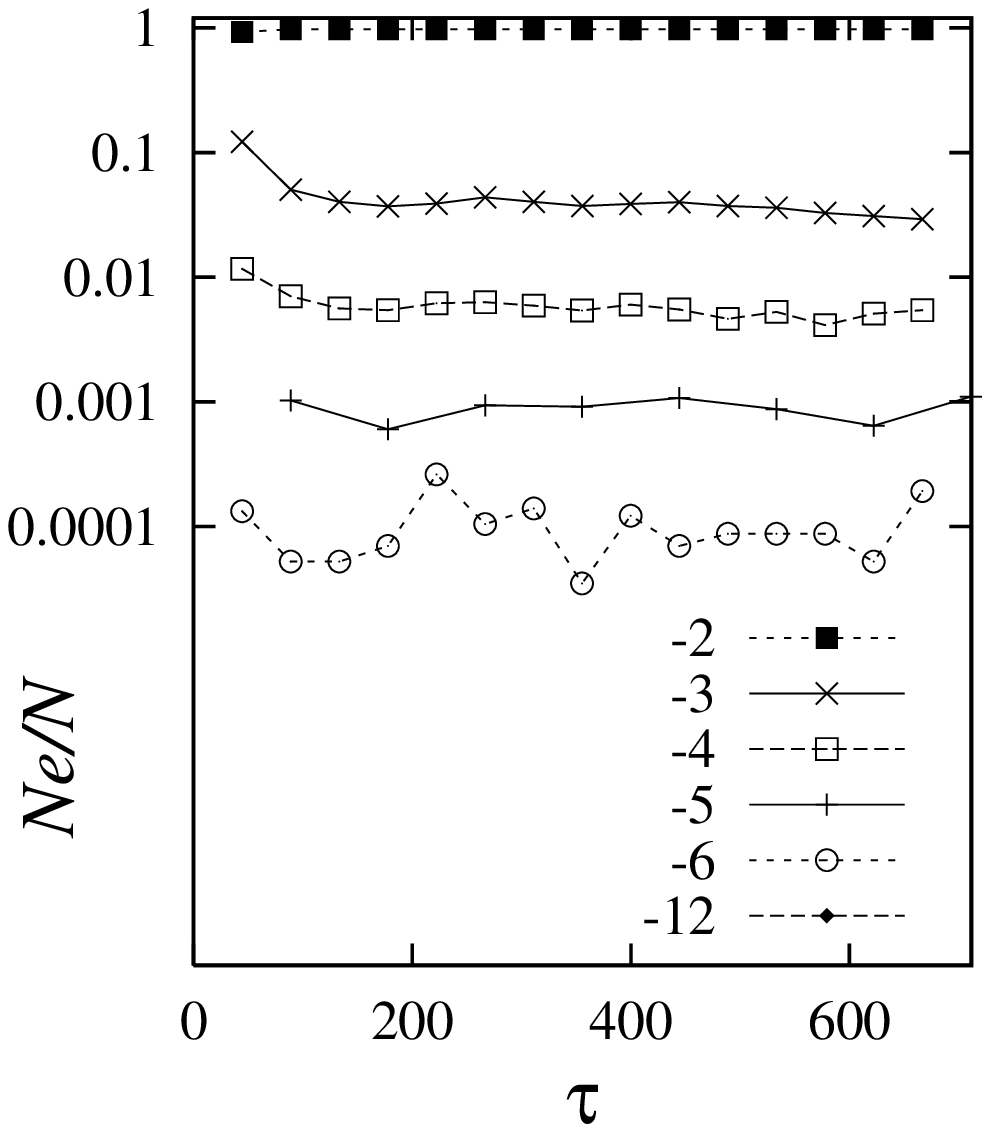,height=5.8cm,angle=-90} \hfill
\epsfig{file=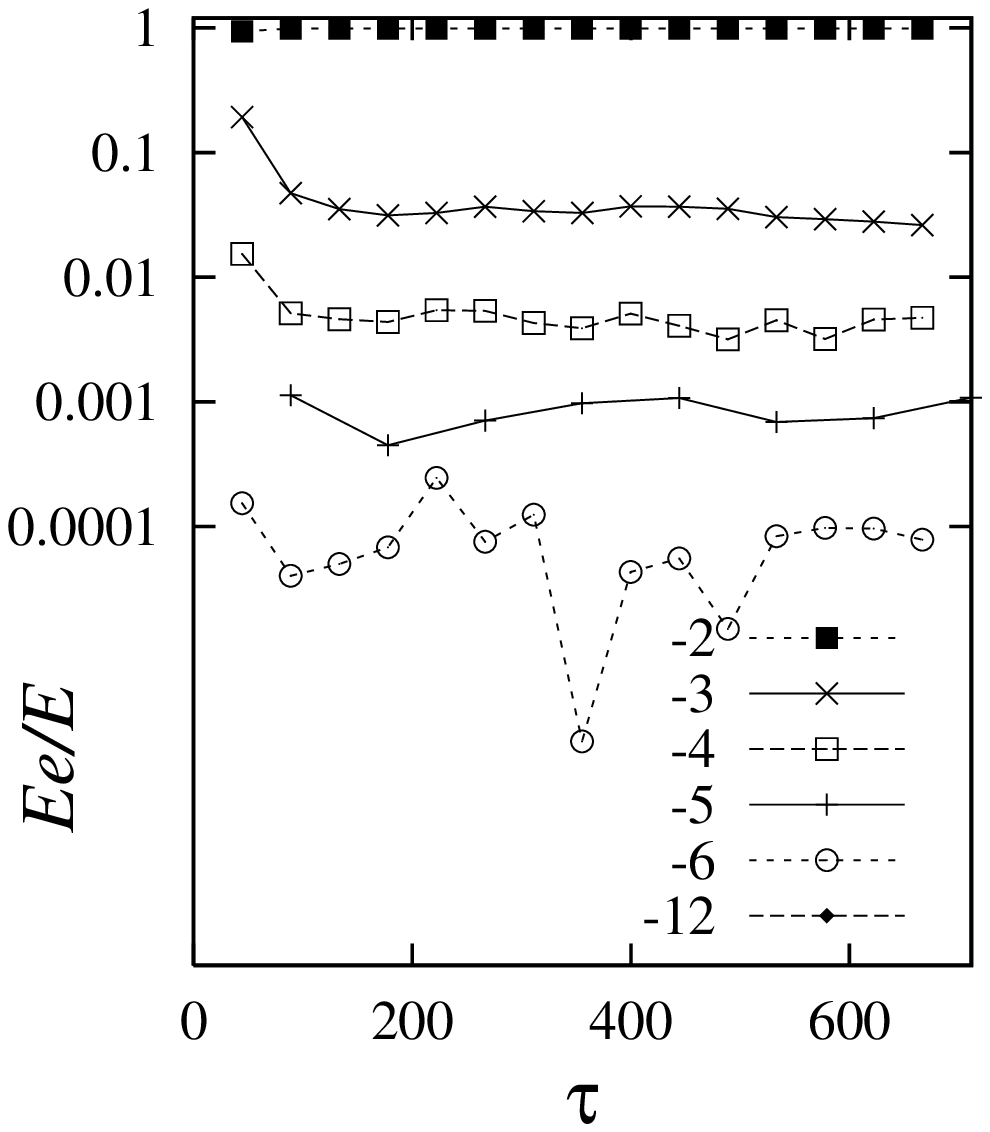,height=5.8cm,angle=-90} \\
\caption{
(a) The fraction of elastic collisions ${\cal C}_e/{\cal C}$ as
function of $\tau$ with $r = 0.4$ and $\log_{10} t_c$ as given in
the inset. The simulations are the same as in Fig.\ \protect\ref{fig:tcfig01a},
however, each point represents an average over all collisions
in the time interval $\Delta t = 0.1$\,s.
(b) The fraction of elastic particles $N_e/N$ plotted against $\tau$
for the same simulations as in (a). Each data point represents an average
over 20 snapshots in the time interval $\Delta t = 0.1$\,s. Note the
different vertical axis scaling in (a) and (b).
(c) The fraction of elastic energy $E_e/E$ plotted against $\tau$
for the same simulations and the same averaging procedure as in (b).
}
\label{fig:tcfig02}
\end{figure*}
The simulations with the largest $t_c$ cool slowest since the number of 
elastic collisions increases with $t_c$, see Eq.\ (\ref{eq:Ce}).
Inserting $t_c = 10^{-3}$\,s and $t_E^{-1}(0)$ into Eq.\ (\ref{eq:Ce}),
which was derived for elastic system in equilibrium however,
leads to $c_e \approx 0.59$, whereas the contact duration of
$t_c=10^{-4}$\,s leads to a much smaller fraction of elastic
contatcs $c_e \approx 0.085$. Evidently, dissipation 
is almost inactive when $t_c$ is as large as $t_c = 10^{-2}$\,s
so that $c_e \approx 0.9999$. On the other hand, the effect of
$t_c$ will be negligible for $t_c < 10^{-5}$\,s when the fraction
of elastic collisions vanishes $c_e < 9 \times 10^{-4}$.
In summary, the contact duration $t_c$ slows down dissipation in the
system -- the larger $t_c$ the stronger the effect. For 
small enough $t_c$ the system is not affected and the variation of
the energy $T$ with the contact duration $t_c$ is rather weak. 
Note that other quantities, as e.g.~the total number of collisions 
per particle $C_t/N$, can vary strongly with $t_c$, as evidenced
in Fig.\ \ref{fig:tcfig01b}. The smaller $t_c$ the more collisions
occur in the system during a fixed time interval. Sometimes, for
$t_c \le 10^{-10}$\,s, jumps in $C_t/N$ are observed. As a consequence,
the total number of collisions looses its meaning as a system-inherent
time scale as soon as the system becomes inhomogeneous. This can also
be understood when recalling that an inhomogeneous system is a system
that has non-constant density, fluctuation velocity, pressure, and 
collision rate. Different parts of the system, evolving with different
collision rates, cannot be assumed to follow a common time-scale.
On the other hand, the variation of $C_t/N$ with $t_c$ allows to see the 
TC model also as a way to reduce the computational effort by decreasing the
global number of events that have to be handled, however, without affecting
physical observables like $T$.\\
Another way to report the effect of the TC model on the simulation
results is to take a closer look at the fractions of elastic 
collisions, particles, and energy in the system. In Fig.\ 
\ref{fig:tcfig02}(a), (b), and (c) these quantities are 
displayed respectively. The fraction of elastic collisions 
$c_e = {\cal C}_e/{\cal C}$ is systematically larger than both
the fraction of elastic particles $n_e=N_e/N$ and the fraction
of elastic energy in the system $e_e=E_e/E$. The latter two 
quantities are always comparable, whereas $c_e$ depends on 
$t_c$ in a different way. 
The fraction of elastic collisions is correlated to $t_c$, at the 
beginning of the simulation, as long as the system is homogeneous.
For larger $\tau$ the fraction of elastic collisions fluctuates
around at a finite value,
independent how small $t_c$ is. Even the simulation with 
$t_c = 10^{-12}$\,s has a rather large fraction of elastic 
collisions. However, setting a large ${\cal C}_e$ in relation to the
large total number of collisions $C_t/N$, or equivalently ${\cal C}_r$,
leads to the conclusion that the TC model adjusts $c_e$ to an
finite, almost constant value. In contrast,
the elastic energy and the fraction of elastic particles
decrease systematically with $t_c$. These two quantities
are directly affected by $t_c$. Concerning averages we must remark
that the procedure to obtain ${\cal C}$ is different from the one 
used to obtain average particle numbers or energies. While {\em all}
collisions in the averaging time interval are summed up to $\cal C$, 
only a certain number of snapshots is 
evaluated to compute $\left < N_e \right >$ and $\left < E_e \right >$. 

\subsection{Variation of the restitution coefficient}

For the parameter study in this subsection, the system with  $N=5740$, 
$\nu=0.227$, and a fixed contact duration $t_c=10^{-5}$\,s is used. 
This choice of $t_c$ is arbitrary, however, as we have shown in the 
previous subsection, it is a reasonable choice in
the framework of the TC model: $t_c$ has a negligible effect on $T$.
Furthermore, such a $t_c$ value is of the
same order of magnitude as the contact duration of e.g.~two steel
spheres with radius $a=1.5$\,mm \cite{luding94d}.

\begin{figure}[htb]
\begin{center}
\epsfig{file=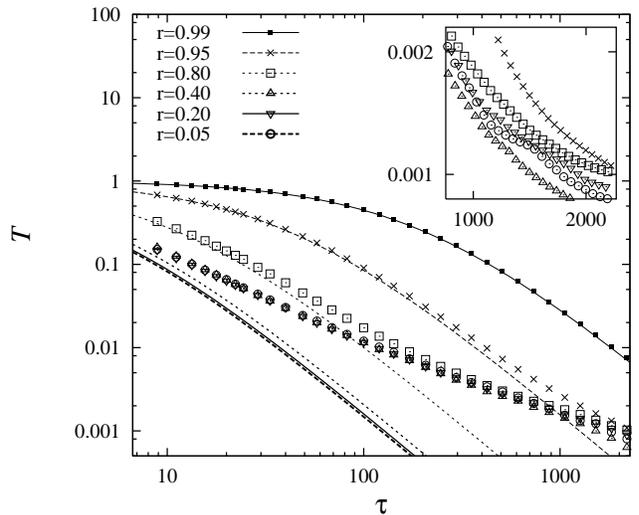,height=8.5cm,angle=-90}
\end{center}
\caption{
Dimensionless energy $T$ plotted against $\tau$ from simulations
with $t_c=10^{-5}$\,s and $r$ as given in the plot. The inset is
a zoom into the plot at large $\tau$ and small $T$. The points are
simulation data, the lines give Eq.\ (\ref{eq:HCS}) for the corresponding
$r$ values.
}
\label{fig:rfig01}
\end{figure}

\begin{figure*}[bht]
~~~~~ {\large (a)} \hfill~\hfill
     {\large (b)} \hfill~\hfill
     {\large (c)} ~~~~~~~~~~~ \hfill ~ \\
\epsfig{file=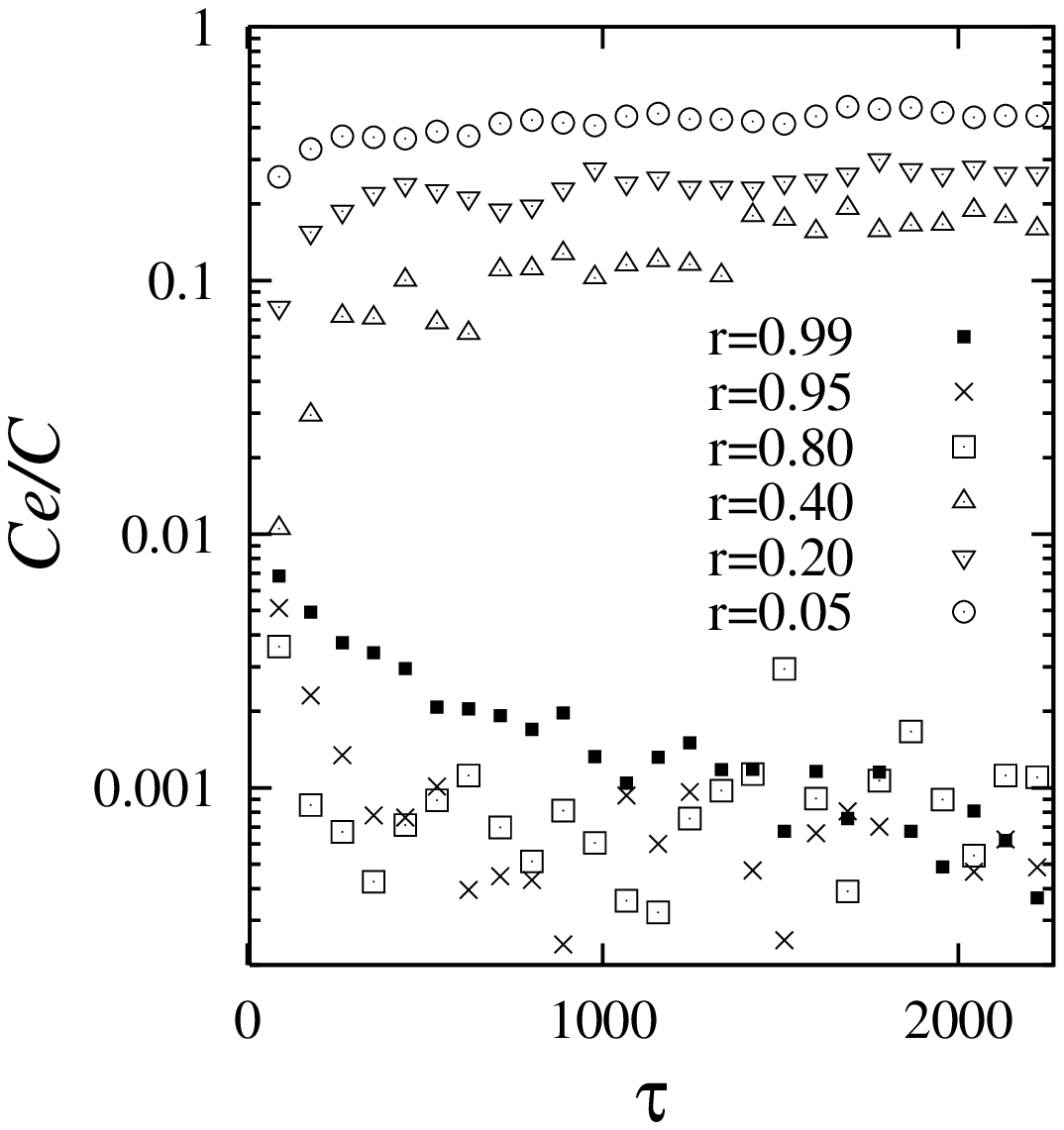,height=5.8cm,angle=-90} \hfill
\epsfig{file=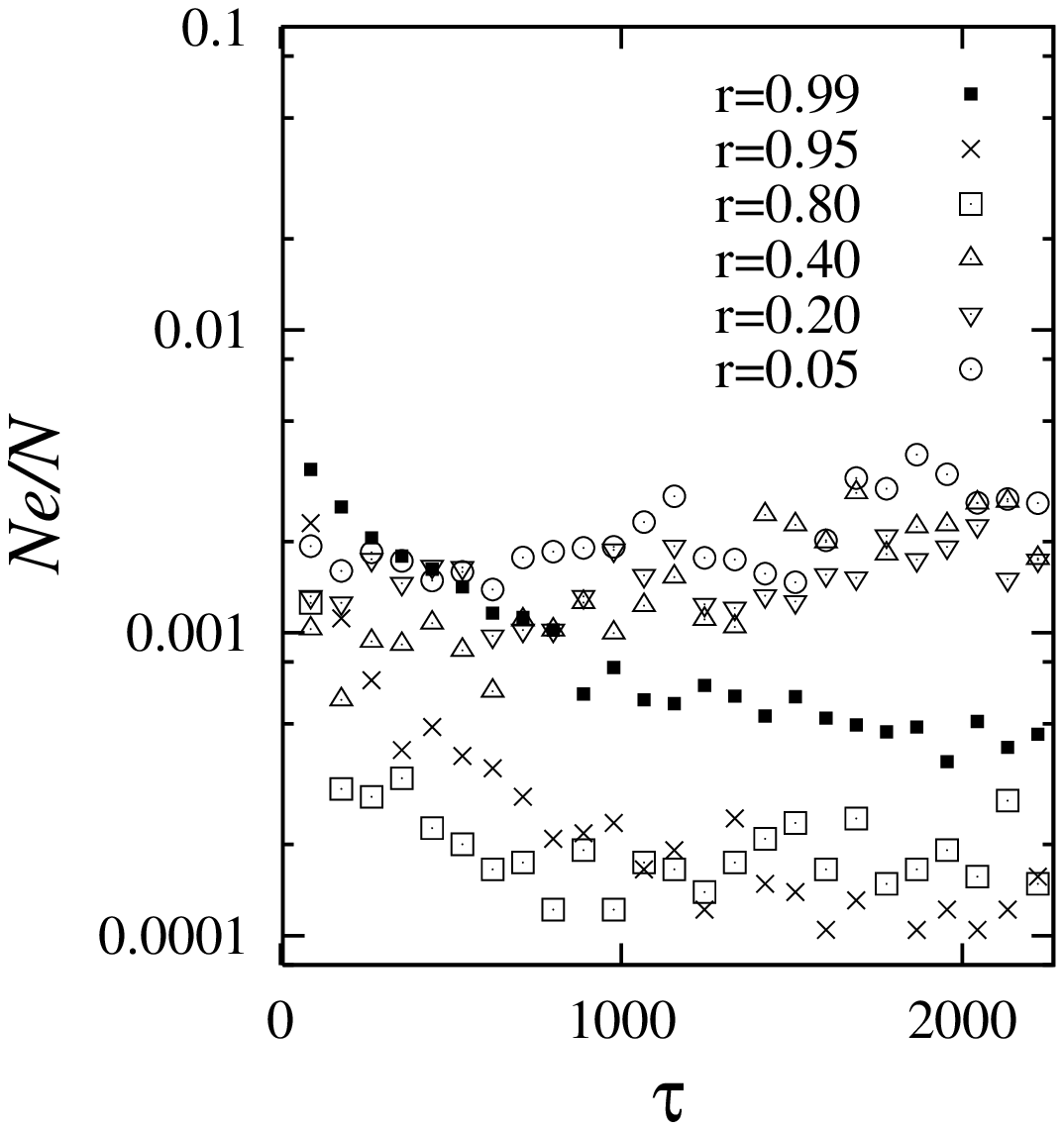,height=5.8cm,angle=-90} \hfill
\epsfig{file=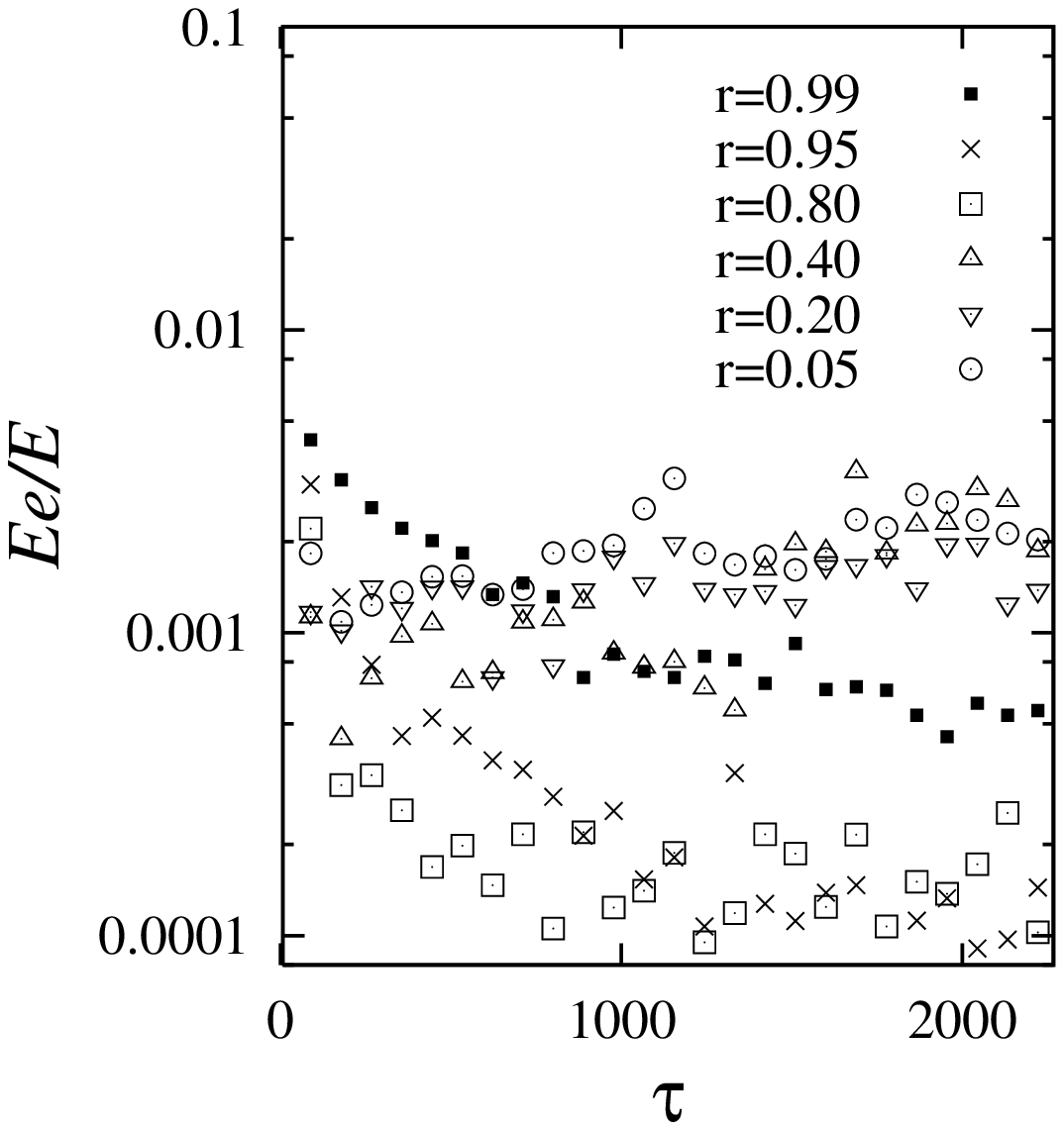,height=5.8cm,angle=-90} \\
\caption{
(a) The fraction of elastic collisions ${\cal C}_e/{\cal C}$ as
function of $\tau$ with $t_c=10^{-5}$\,s and $r$ as given in
the inset. The simulations are the same as in Fig.\ \protect\ref{fig:rfig01},
however, each point represents an average over all collisions
in the time interval $\Delta t = 0.2$\,s.
(b) The fraction of elastic particles $N_e/N$ plotted against $\tau$
for the same simulations as in (a). Each data point represents an average
over 20 snapshots in the time interval $\Delta t = 0.2$\,s. Note the
different vertical axis scaling in (a) and (b).
(c) The fraction of elastic energy $E_e/E$ plotted against $\tau$
for the same simulations and the same averaging procedure as in (b).
}
\label{fig:rfig02}
\end{figure*}

In Fig.\ \ref{fig:rfig01}, the dimensionless kinetic energy 
$T$ is again plotted against the rescaled time 
$\tau$, with the initial collision 
rate $t_E^{-1}(0) \approx 444$\,s$^{-1}$. Simulations are
performed for different restitution coefficients as given in
the plot, and are compared to the analytical solution of the
homogeneous cooling state, see Eq.\ (\ref{eq:HCS}). Only for
$r=0.99$, we evidence reasonable agreement with the theory.
With decreasing $r$, deviations from the theoretical curve occur
already for smaller and smaller $\tau$. However, for $r \le 0.4$, there
are only small differences between simulations with different $r$, 
as can be seen more clearly in the inset.

We do not present a plot of $T$ against $C_t/N$ here, but make
some qualitative remarks on the total number of collisions.
We evidence that the total number of collisions per particle varies 
with $r$ as it varies with $t_c$. For large $r$ and small $C_t/N$ 
the energy behaves as $T = \exp[- \frac{1}{2}(1-r^2) C_t/N]$, as can 
be simply derived using Eq.\ (\ref{eq:HCS}) and integrating the
collision rate ${\cal C}_r$ over $\tau$. This behavior is obtained only
for $r \approx 1$, already for $r \le 0.95$ the simulations 
deviate from the theoretical prediction.

For strong dissipation, only {\em a few} particles perform 
many collisions. One can expect that the time between the
collisions of such particles drops below the threshold $t_c$ so that 
$r$ is set to unity according to Eq.\ (\ref{eq:epsnew}). The TC model is 
active, hinders dissipation, and thus evades the inelastic collapse.
The number of particles affected by the TC model will be discussed in the
following.
In Fig.\ \ref{fig:rfig02} we present data on $c_e$, $n_e$, and
$e_e$ in a way similar to Fig.\ \ref{fig:tcfig02}. Again, we
obtain that $c_e$ behaves differently from $n_e$ and $e_e$.
All simulations with $r \ge 0.80$ have a negligible fraction 
of elastic collisions ($c_e < 2 \times 10^{-3}$ for long times).
The simulations with stronger dissipation ($r < 0.80$) have 
an increasing number of elastic collisions with increasing
dissipation, i.e.~decreasing $r$.

From the data on $N_e$, we can estimate the number of particles
with the largest collision rate, which typically are involved into 
elastic collisons. A fraction of
$n_e=0.002$ corresponds in the case of $N=5740$ to $N_e \approx 10$
particles. Even in the case of very strong dissipation only 
about 10 particles are affected by the TC model.

\section{Conclusion and Outlook}

In this study we discussed the TC model, an extension of the frequently 
used inelastic hard-sphere model. Introducing the contact duration $t_c$
as a material parameter, multiparticle interactions are defined in some 
sense. They concern particles with large collision rates which are 
assumed to contribute to the elastic energy in the system 
(which cannot be dissipated). The TC model can reach a quasi-static
situation when only elastic and potential energy remain.
Dissipation is locally inactive for large collision rates, 
i.e.~the elastic limit where multiparticle contacts occur, 
and active for rare events, i.e.~the dissipative limit where contacts
are binary almost always. 
The TC model allows simulations in ranges of parameter space, where the 
classical inelastic hard-sphere model breaks down due to the inelastic 
collapse. Potential, kinetic, and elastic energies as well as stresses 
and forces are defined as averages over time-intervals comparable to $t_c$.
The material parameter $t_c$ can be identified with the contact
duration $t_c^{el}$ of a simple linear particle model, involving 
e.g.~particle mass and stiffness. 

Furthermore, mean-field estimates for the fraction of elastic particles 
$n_e$, the fraction of elastic collisions $c_e$, and the fraction of 
elastic energy $e_e$ in the system are presented. 
The simulation results indicate that a more elaborate theory
is required to explain the obtained discrepancies. However, 
the definition of $n_e$, $c_e$, and $e_e$ is also valid 
in non-equilibrium, dissipative systems. 
Detailed examinations of the inhomogeneously cooling situation
leads to the conclusion that the TC model affects a small fraction
of the particles only -- just as the inelastic collapse. Therefore,
we beleive that the TC model removes inelastic collapse in a 
physically reasonable way, without disturbing the global behavior
of the system, i.e.~the clustering. This is strictly true for
realistically small $t_c$ values, because an extremely large $t_c$ 
changes the global behavior dramatically.

The TC model is defined for arbitrary dimension so that an 
extension to three dimensional systems is straightforwardly
performed \cite{luding98d}, and also gravity or moving walls
can be implemented \cite{luding96e,luding97c} without loosing
its generality. A future aim is to include the TC model into 
kinetic theories in the style of Haff \cite{haff83} where
it will affect only the energy dissipation rate via a 
correction factor $1-c_e$. However,
besides excluded volume and dissipation, one important property of
granular materials, namely friction, is missing in its present form.
A proper definition of friction in the framework of 
the TC model is in progress and has to be tested with systems like
particles on an inclined plane or sandpiles which are kept at rest by 
friction. 


\end{document}